\documentclass[letterpaper,11pt]{article}
\usepackage{jheppub_mod}
\usepackage{graphicx, amsmath, amssymb, amstext}
\usepackage{relsize, float, multirow}
\usepackage{array, epstopdf, hyperref, }
\usepackage{caption, subcaption, mathtools}
\usepackage{newtxmath}

\title{CMB spectral distortions constraints on  primordial black holes,
  cosmic strings and long lived unstable particles revisited}
\author[a]{Sandeep Kumar Acharya,} 
\author[a]{Rishi Khatri}
\affiliation[a]{Department of Theoretical Physics, Tata Institute of 
Fundamental Research, Mumbai 400005, India}
\emailAdd{sandeepkumar@theory.tifr.res.in, khatri@theory.tifr.res.in}
\date{\today}
\abstract{ 
We  calculate the spectral distortions from Hawking evaporation of
primordial black holes before the epoch of recombination, taking into account emission of all standard model
particles, including quark and gluons, and evolving the resulting particle
cascades in the expanding Universe. We show that the constraints on the
abundance of primordial black holes are
stronger by more than an order of magnitude compared to the previous
calculations which take only the primary photon emission into account.  We
also show that the shapes of the spectral distortions is different from the
$y$ or $i$-type distortions and are sensitive to  the mass of the
primordial black holes. We also extend previous  constraints on the decay
of long lived unstable particles before recombination to additional
decay channels.  We show that for dark matter mass $\lesssim$ 1 GeV,
the spectral distortion shape is a function of the dark matter mass as well
as the decay channel to standard model particles. We also provide new
spectral distortion constraints on superconducting cosmic string decay. We
explicitly show that consideration of emitted photon spectrum from string
decay is not only important for the future experiments but also for already
available COBE-FIRAS data.      
}
\notoc
\begin{document}
\maketitle
\section{\label{sec:intro}Introduction}
\hspace{1cm}Spectral distortions of the cosmic microwave background (CMB),
abundance of light elements produced in the big bang nucleosynthesis (BBN),
and CMB anisotropy power spectrum provide multiple avenues to constrain
external electromagnetic energy injection into the primordial baryon-photon
fluid in the early universe. Injection of energy around the recombination
era ($z\sim$1000) modifies the standard recombination history, damping the
CMB temperature anisotropy power spectrum while boosting the polarization
power spectrum \cite{ASS1998,Chen:2003gz,Galli:2013dna}. Recently, it was
shown that the CMB anisotropies can constrain energy injection at redshifts
as high as $z\sim$ 10000 \cite{AK20192}. CMB temperature anisotropy power
spectrum has been measured to exquisite precision by Planck \cite{Pl2018}
and ground based experiments such as SPT \cite{SPT2016}, and ACT
\cite{ACT2017}. Next generation experiments
\cite{Hazumi2019,Pixie2011,CORE2018,PICO2019,CMBS42019}, with more precise
polarisation measurement, can strengthen the present day constraints by an
order of magnitude. Energetic photons in the electromagnetic cascades,
triggered by the injection of high energy particles, can photo-dissociate
light elements produced by BBN, changing their abundances. BBN can
constrain energy injection upto very high redshifts
\cite{PS2015,HSW2018,FMW2019}. BBN constraints, limited by the
astrophysical uncertainties on the measurement of the abundance of light
elements, are well established
\cite{ENS1985,EGLNS1992,KM1995,KKMT2018,AK20192} and
unlikely to improve significantly in the near future. The importance of
consistent evolution of electromagnetic cascades when calculating BBN
constraints was emphasized recently in \cite{AK20192}. \par
\hspace{1cm}
The best measurement of the CMB spectrum was done more than 25 years ago by COBE-FIRAS (Cosmic Background Explorer- Far InfraRed Absolute Spectrophotometer) \cite{F1996,FM2002}.
In contrast to the CMB anisotropies and the BBN, the spectral distortion
constraints have the potential for many orders of magnitude improvement
with near future  experiments like PIXIE \cite{Pixie2011}. At $z\gtrsim
2\times 10^6$, photon non-conserving processes such as double Compton
scattering and bremsstrahlung together with Compton scattering, establish a
Planck spectrum and exponentially suppress any departure from
equilibrium. Any energy injection above $z\gtrsim 2\times 10^6$ in the form
of heating, therefore, raises the temperature of the photons while
preserving the blackbody spectrum
\cite{Sz19701,dd1982,Chluba:2011hw,ks2012}. At $z\lesssim 2\times 10^6$,
energy injection can give rise to $y$, $\mu$ or $i$-type distortions
\cite{Sz1969,Sz19701,Is19752,Bdd1991,Chluba:2011hw,Ks2012b,Chluba:2013vsa}.
Previous calculations assumed that all of the injected energy goes into
heating up the background electrons which boost the CMB photons by
non-relativistic Compton scattering. Recently, it was shown that for high
energy ($>$10 keV) electron-positron and photon injections, significant
relativistic corrections are expected \cite{AK2018}. Relativistic electrons
and positrons produced during the particle cascade, triggered by the high
energy injected particle, can boost the CMB photons to significantly higher
energy compared to the $y$-type distortion, imprinting characteristic
energy dependence on spectral distortion shapes due to relativistic
collision processes. These ultra-relativistic distortions modify the constraints on energy injections. It was shown in
\cite{AK2019} that for dark matter decay to monochromatic electron-positron
pairs or photon pairs, the constraints can relax by upto a factor of 4-5
for injection redshifts $z_{\rm inj}<$50000. In this paper, we extend these calculations to additional dark matter decay channels, Hawking evaporation of primordial black holes and decay of cosmic strings. \par
\hspace{1cm}    
There are many particle physics models beyond the standard model which can
provide a dark matter candidate. The simplest candidate  is the WIMP (Weakly Interacting Massive Particle)  which
invokes just one extra particle with weak interaction to explain the
present density of dark matter. However, in general, in theories such as
supersymmetric and Kaluza-Klein type extra-dimensional models, there are 
large number of new particles with the lightest stable particle being the
dark matter candidate. A detailed discussion of possible dark matter
candidates can be found in reviews such as \cite{BHS2005,F2010}. These
models are invoked not only to explain dark matter but also to explain
problems related to standard model physics, for example, the hierarchy
problem. These models typically include small coupling with standard model
particles so as not to violate the collider and astrophysics constraints
while simultaneously alleviating problems related to particle
physics. In these models, it is perfectly reasonable to assume that many new
particles were produced  in the early Universe. The lightest stable
particle is the dark matter today. However, there could have been
additional particles which were unstable and decayed later into standard
model particles and/or dark matter particles. If the lifetime of these new
particles is larger than $\sim 2$ months (decays happening at $z\lesssim 2\times 10^6$), then a fraction of energy going into the standard model particles will give rise to CMB spectral distortions.  \par
  \hspace{1cm}   
  Another interesting scenario, which can inject energetic electromagnetic
  particles at high redshifts, is Hawking evaporation of primordial black
  holes and accreting primordial black holes. We will only
  consider evaporting primordial black holes in this paper, ignoring
  accretion. Accretion can release additional energy and make the
  constraints stronger, especially towards the higher mass end, albeit with
  significant uncertainties related to  the modeling \cite{ROM2008,Am2017,PSCCK2017}.  Evaporating black holes emit energetic standard model particles
  having energies of the order of the temperature of the black hole which
  in turn depends on the mass of the black hole \cite{H1974,H1975}. The
  shape of the spectrum of emitted particles at energies much  larger
  compared to the black hole temperature is thermal. However, at low and
  intermediate energies, there can be significant deviation from the
  thermal spectrum \cite{P1976,MW1990}. Black holes with temperature
  $\gtrsim$ 1 GeV emit mostly quarks and gluons which produce significant
  amount of secondary electrons, positrons and photons after hadronization
  \cite{MW1991}. Spectral distortion constraints for primordial black holes
  with mass range $\sim10^{11}$g - $10^{13}$g (corresponding to black hole
  evaporation between y and $\mu$ era) have previously been studied in
  \cite{TS2008}. However, they  only considered the primary photon emission
   from the black holes and ignored   emission  of other particles. 
  Recent work of \cite{LSHLC2019} takes into account all particles but
  assumes that all energy goes into non-relativistic $y,i$ and $\mu$-type
  distortions (referred collectively hereafter as the $yim$
  distortions). Black holes emit all particles democratically provided  the
  emission is kinematically allowed  i.e. the temperature of the
  evaporating black hole is of the same order or larger compared to the mass of
  the emitted particle. The fractions of energy going into different
  particles are proportional to the multiplicity of each particle. For
  example, 6 quarks with 3 colors, antiparticles and spins of $1/2$ have multiplicity
  72 and account for the bulk of emitted energy while the  photons with
  multiplicity of 2 corresponding to 2 polarizations account for a
  sub-dominant fraction of evaporating energy. Therefore, it is important to take into account secondary photon, electron, and positron emission from quarks and gluons at high temperatures where primary photon emission only accounts for a tiny fraction of the total emitted energy. In this work, we take into account all the particles emitted by black holes and calculate the non-thermal ($ntr$) spectral distortion shape by evolving the electromagnetic cascades. We compare the full non-thermal distortion spectrum with thermal distortion spectrum (${yim}$ distortion). For black hole evaporation at $z\lesssim 10^5$, the shape of non-thermal distortion is significantly different from that of thermal non-relativistic spectral distortions, while for higher redshifts, spectral distortions are thermalized to $\mu$ distortion. \par
  \hspace{1cm}
Phase transitions in the early universe can give rise to topological
defects like monopoles, strings and domain walls \cite{K1976}. String
theory also predicts the existence of such objects \cite{CK2010}. These
cosmic strings carry energy, source perturbations in surrounding matter,
and thus, can have observable effects in the CMB anisotropy power spectrum
\cite{BPRS2002,MP2014}. Cosmic strings can be superconducting and can decay
by emitting their energy as high energy electromagnetic particles
\cite{W1985} and can therefore create CMB spectral
  distortions. Since the spectrum of photons emitted by the superconducting
  cosmic strings is highly non-thermal and depends on the cosmic string tension
  and current, we expect that the shape of the resulting CMB spectral distortions to also be
  different from the \emph{yim} distortions and in particular be sensitive
  to the parameters of the cosmic string.
   \par
  \hspace{1cm}   
   We use the COBE-FIRAS \cite{F1996} data to give constraints for different
   energy injection scenarios using CMB spectral distortions calculated by
   evolving the electromagnetic cascades. In particular, we do not use the
   $yim$ approximation but use the actual spectral distortion shapes. We
   also give forecasts for a future PIXIE-like experiment assuming a factor
   of 1000 enhancement in the sensitivity over COBE-FIRAS.
 We use Planck \cite{Pl2018} $\Lambda$CDM
 cosmological parameters for all calculations.
 \section{\label{sec:cascade}Electromagnetic cascades in the expanding Universe}
  In this section we briefly explain the qualitative aspects of
  electromagnetic cascade evolution. The technical details can be found in
  \cite{AK2018}. We follow the computational approach of
  \cite{Slatyer:2009yq,Kanzaki:2008qb,Kanzaki:2009hf} in an ionized
  universe as described in \cite{AK2018}. We divide the energy
 range of interest into 500 logarithmically-spaced energy bins for each particle (e.g. photons,
 electrons and positrons) in dimensionless frequency variable
 $x=E/T_{\mathrm{CMB}}$, where $E$ is the kinetic energy of the electron or
 positron or the energy of the photon, and $T=2.725(1+z)\,$K is the CMB
 temperature. A high energy injected
 particle, after repeated scattering, deposits its energy by boosting the
 background particles. In an ionized universe, electrons with energy below keV
 deposit their energy via heat while higher energy electrons boost the CMB
 photons by inverse Compton scattering process (ICS). Positrons can release
 two photons with energy 511 keV after annihilating with the background
 electrons. The rate of energy deposition for electrons and positrons is much faster  compared to the Hubble
   rate  (Fig. 2 of \cite{AK2018}) and it is a very good
   approximation to assume that the electrons and positrons deposit their
   energy instantaneously.  
Photons on the other hand deposit their energy at a comparable or slower
rate compared to the expansion rate and the photon spectrum needs to be evolved
by taking the background expansion into account. We take into account collision processes such as Compton scattering, photon-photon elastic
 scattering and electron-positron pair production. 
 We are essentially solving a system of coupled differential equations (Eq. 3.1 of
 \cite{AK2018}) where the  particles in the cascade move from one energy bin to
 another as the cascade evolves. The  probabilities or rates of transfer
 between the energy bins  are determined by the competition between the different
 collision processes and the expansion rate of the universe. These particles in the cascade have
 relativistic energies with non-thermal distribution. The shape of the CMB spectral
 distortions  from these non-thermal particles is significantly different 
 from the $y$-type and the $i$-type distortions. For
 non-thermal photons, with energy much higher compared to the CMB photons,
 we can ignore the energy distribution of the background electrons. For
 photons with energy comparable to the CMB photons, we take into account
 Doppler broadening due to the thermal motion of the background
 electrons. We also take into account  stimulated Compton scattering at photon
 energies comparable to the background CMB photons. At  $z\gtrsim
 10^5$, the non-thermal relativistic CMB spectral distortions also thermalize to $\mu$-type distortion
 \cite{AK2019} and become a function of only the total injected energy,
 losing sensitivity to the initial spectrum of injected particles. 
     
\section{\label{sec:blackhole}Primordial black hole evaporation}
Primordial black holes are formed from high energy density peaks in the
early universe when overdense regions decouple from the Hubble expansion
and collapse under their self-gravity \cite{ZN1966,H1971,CH1974}. If the density
perturbation at the stage of horizon entry exceeds a threshold, determined
by the thermal pressure, it can lead to the formation of a black hole with mass
of the order of horizon mass, $M_{\rm BH}$ $\sim\frac{c^3t}{G}$
\cite{CKSY2010}, where t is the cosmological time, G is Newton's
gravitational constant and c is the speed of light. Depending on the
formation epoch, primordial black holes can span huge range of mass, from
Planck scale to hundreds of solar mass.  We will
consider only Schwarzschild black holes in this paper. This is a good
approximation because for a rotating black hole more than 50 percent of black hole energy is emitted when the black hole has already lost most of its spin and  is slowly rotating \cite{P19761}.

Black holes, once formed, radiate particles as a hot body with temperature (with $c=\hbar=k_B$=1) \cite{H1974,CKSY2010},
\begin{equation}
T_{\rm BH}=\frac{1}{8\pi GM_{\rm BH}}=1.06M^{-1}_{\mathrm{10}} \, \rm{TeV}
\end{equation}
where $M_{10}=M_{\rm BH}/10^{10}g$ is the mass of black hole in units of $10^{10}$ g. \par       
 \hspace{1cm}
 The lifetime of evaporating black holes is approximately given by \cite{MW1991,CKSY2010},
 \begin{equation}
 t_{\rm BH}=407\left[\frac{f(M_{\rm BH})}{15.35}\right]^{-1}M^3_{10}\,{\rm s},
 \end{equation}
 where $f(M_{\rm BH})$ carries the information about the emitted particles
 and is defined below.
 The particle emission rate from the Schwarzschild black holes in the energy interval E and E+dE is given by \cite{P1976,MW1990}, 
 \begin{equation}
\frac{dN}{dt}=\frac{\Gamma_s(E,M_{\rm BH},m)}{2\pi}\frac{1}{e^{E/T_{\rm BH}}-(-1)^{2s}}dE,
\end{equation}  
where $\Gamma_s$ is the absorption coefficient which depends on the spin
($s$) and mass ($m$) of the emitted particle and the mass of the black hole
\cite{P1976,MW1990}. The spectrum of emitted particles just outside the
black hole horizon has a thermal distribution. These particles experience
an effective potential outside the horizon which is determined by the mass
of black hole and the angular momentum and mass of the emitted particle
\cite{RW1957}. Therefore, the emission spectrum of the particles far way
from the horizon becomes mass- and spin-dependent. These absorption coefficients become mass (of particle) and  spin-independent at high energy  approaching the limit, $\Gamma_s(E/T_{\rm BH}>>1)=27G^2M^2$. The PBH mass loss rate can be written as \cite{MW1991}, 
\begin{equation}
\frac{dM_{\rm BH}}{dt}=-5.34\times 10^{25} \newline
\frac{f(M_{\rm BH})}{M^2_{\rm BH}}\,{\rm  g s^{-1}}.
\end{equation} 
where $f(M_{\rm BH})$ is approximately given by,
\begin{multline}
f(M_{\rm BH})=1.569+0.569[\exp(-\frac{M_{\rm BH}}{x_{{\rm bh},s=1/2} M^T_{\mu}})+
3\exp(-\frac{M_{\rm BH}}{x_{{\rm bh},s=1/2} M^T_{u}})+3\exp(-\frac{M_{\rm BH}}{x_{{\rm bh},s=1/2} M^d_{d}})\\ +3\exp(-\frac{M_{\rm BH}}{x_{{\rm bh},s=1/2} M^T_{s}})+ 
3\exp(-\frac{M_{\rm BH}}{x_{{\rm bh},s=1/2} M^T_{c}})+\exp(-\frac{M_{\rm BH}}{x_{{\rm bh},s=1/2} M^T_{\tau}})+3\exp(-\frac{M_{\rm BH}}{x_{{\rm bh},s=1/2} M^T_{b}})\\
+3\exp(-\frac{M_{\rm BH}}{x_{{\rm bh},s=1/2} M^T_{t}})]+0.963\exp(-\frac{M_{\rm BH}}{x_{{\rm bh},s=1} M^T_{g}}),
\label{particle fraction}
\end{multline}
with
\begin{equation}
M^T_i=\frac{1}{8\pi Gm_i},
\end{equation}
 where $M^T_i$'s are the mass of the black holes whose temperature is equal
 to the mass $m_i$ of the respective standard model particle labeled by $i$, $i\in
 {(\mu,u,d,s,c,\tau,b,t,g)}$ for muon, up quark,
 down quark, strange quark, charm quark, tau, bottom quark, top quark and
 gluon  respectively, $x_{\rm bh}=\frac{E}{T_{\rm BH}}$, and $x_{{\rm bh},s}$ is
 the location of the peak of the emitted instantaneous power spectrum i.e the
 dimensionless energy $x_{\rm bh}$ of the peak of the distribution,
 $\psi_s(x_{\rm bh})=\frac{\gamma_s(x_{\rm bh})x^3_{\rm BH}}{e^{x_{\rm
       bh}}-(-1)^{2s}}$ with $\gamma_s(x_{\rm bh})=
 \Gamma_s(E)/27G^2M^2$. The value of $x_{{\rm bh},s}$ for s=0,1/2, and 1 is
 2.66,4.53, and 6.04 respectively \cite{MW1991}. Thus, a standard model particle
 is emitted from a black hole in appreciable amount when the temperature of
 the black hole is such that the peak of the instantaneous power emission from
 the black hole is of the order of the mass  of that standard model particle
 i.e. $x_{{\rm bh},s}\sim \frac{m_i}{T_{\rm BH}} \Rightarrow x_{{\rm bh},s}
 M^T_{i} \sim M_{\rm BH}$. For massless particles or
 for particles with mass much smaller compared to the black hole
 temperature, the contribution to particle emission is given by the
 integral of $\psi_s(x_{\rm bh})$ over all energy, which is 6.89,3.68 and
 1.56 for s=0,1/2 and 1 particles respectively \cite{MAC2016}. The
 normalization of $f(M_{\rm BH})$ is chosen such that the contribution of
 massless particles i.e photons and neutrinos add upto 1. In
 eq. \ref{particle fraction}, we have assumed that photons, electrons,
 positrons and neutrinos are freely emitted which holds true for the
 calculations in this paper, since we consider black holes with $T_{\rm
   BH}>m_e$. The contribution from photons  $(2\times 0.06)$, neutrinos
 ($6\times 0.147$) and $e^-,e^+$ ($4\times 0.142$), adds upto $1.569$. \par
\hspace{1cm}
We can  see from Eq. \ref{particle fraction} that quarks and gluons dominate the emission spectra when temperature of black hole is within an order of magnitude of their mass. Almost half of their energy is converted to photons, electrons and positrons while the other half is mostly converted to neutrinos (see Fig. 4 of \cite{DM2010}). These secondary electromagnetic particles have to be taken into account when calculating the spectral distortion signals.
\subsection{Calculations and results}
We  us  the absorption
coefficient $\Gamma_s(E,s,M_{\rm BH})$ provided in  \cite{MAC2016} to obtain the spectrum of primary particles emitted from the
black hole. We have
taken into account the change in the mass of the black hole (and therefore
its evolving temperature and emitted instantaneous power) as it
evaporates. To take into account the secondary emission from primary
particles directly emitted from the black hole, we use the tabulated result
of \cite{DM2010} with electroweak corrections. The authors of \cite{DM2010}
have provided fluxes of  secondary products after hadronization ($\gamma$, $e^-e^+$,
neutrinos,  hadrons) as a function of energy for injection of
 different  standard model particles  for initial energies of the particles
 between  5 GeV and 100 TeV using PYTHIA
\cite{PY2008}. We include the decay of  unstable hadrons like pions and neutron   to
stable hadrons, which happens on time scales much smaller compared to the
Hubble time in the redshift range of interest to us, in our
calculation. For black hole evaporation, the branching fraction to
different standard model particles is given by Eq. \ref{particle
  fraction}. We superimpose the secondary flux from all
standard model channels to obtain the total secondary flux from black hole
evaporation. We ignore the energy going into  neutrinos and stable
hadrons. This is a good approximation since the neutrinos only interact
weakly with the baryons and photons and deposit negligible energy.  The error we make by ignoring the energy
going into protons and antiprotons is $\lesssim 15 \%$. We leave the
inclusion of proton and antiproton interactions with the baryon-photon
plasma to future work. For black holes with mass $M_{\rm BH}\approx
10^{13}$g, which decay in the y-era (at redshift $\sim$ 7000), the
temperature of the black hole is $\approx$ 1 GeV and  the average energy of
 quarks and gluons that are emitted is of the order of few GeV. For black
 holes decaying at earlier redshifts (i.e. lower black hole mass), the
 temperature of the black hole and the energy of the emitted particles is even
 higher. The tabulated data in \cite{DM2010} is therefore sufficient for
 the black holes considered in this paper. 
  \par
\hspace{1cm}             
\begin{figure}
\centering
\includegraphics[scale=1.0]{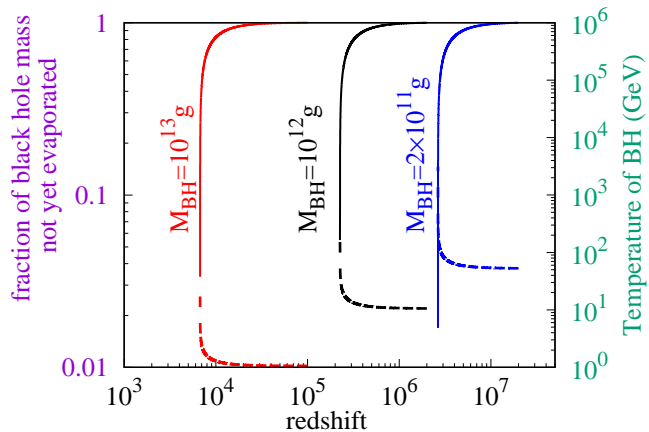}
\caption{Fraction of mass of black holes yet to be evaporated (solid lines) and their temperature (dashed line) as a function of redshift.}
\label{fig:blackholefraction}
\end{figure}
\begin{figure}
\centering
\includegraphics[scale=1.0]{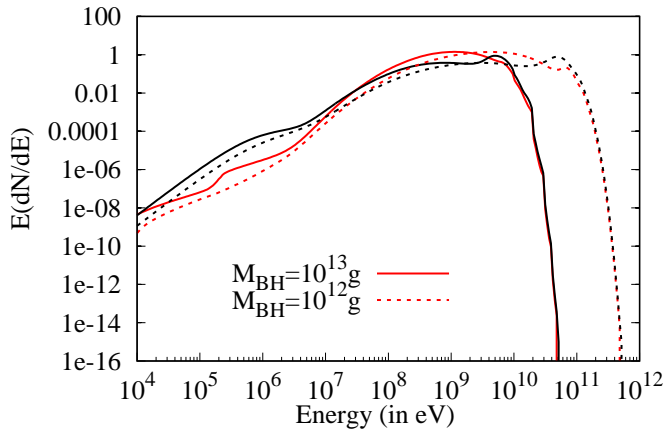}
\caption{Instantaneous spectrum  of photons
  (red) and electron-positron(black) pairs with number of particles per energy (dN/dE) from evaporating primordial black holes as a function of energy at redshift $z=2\times 10^5$ for
  two different mass of black holes. At $z=2\times 10^5$, $10^{13}$g mass black holes are mostly intact. All contributions, primary as well as secondary
  particles after hadronization and decay of unstable particles, are included.}
\label{fig:blackholespectrum}
\end{figure}
\begin{figure}[!tbp]
  \begin{subfigure}[b]{0.4\textwidth}
    \includegraphics[scale=0.8]{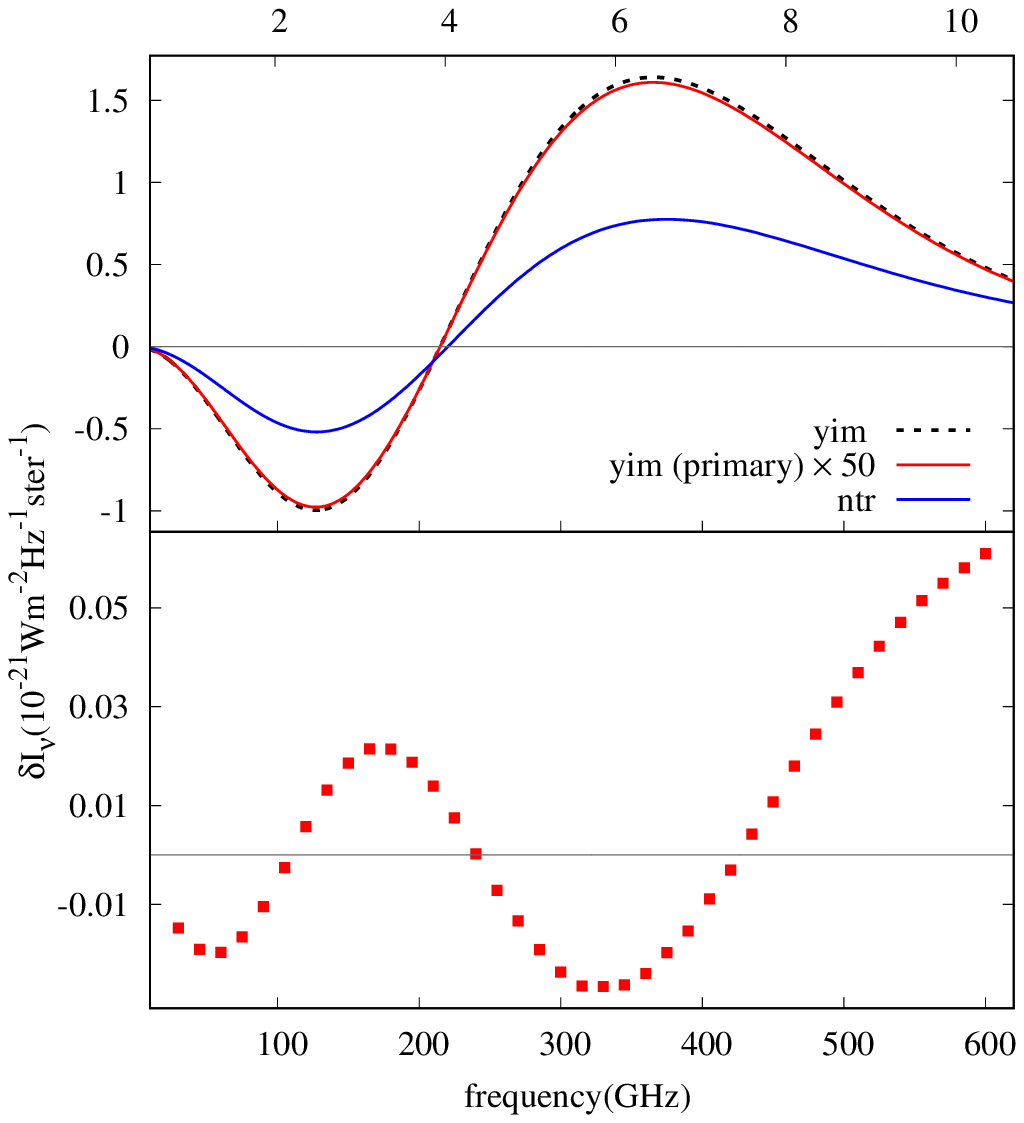}
    \caption{}
    \label{fig:bhevaporation10^13g}
  \end{subfigure}
  \hfill
  \begin{subfigure}[b]{0.4\textwidth}
    \includegraphics[scale=0.8]{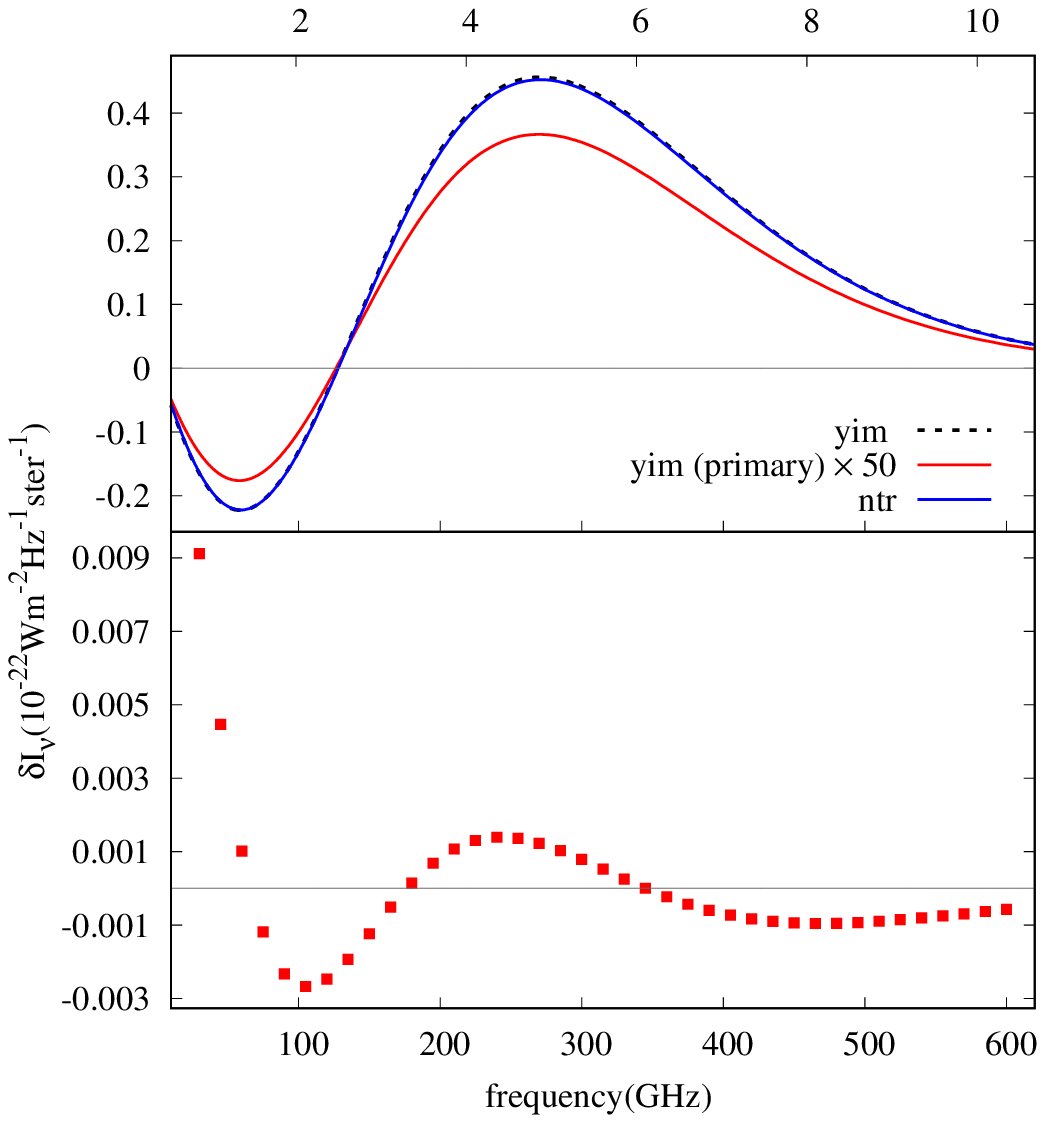}
    \caption{}
    \label{fig:bhevaporation10^12g}
  \end{subfigure}
  \caption{Spectral distortions from evaporating primordial black holes
    with mass $M_{\rm BH}=10^{13}$g (left) and $M_{\rm BH}=10^{12}$g
    (right) for  $f_{PBH}=10^{-3}$. The bottom panel shows the difference
    between the $ntr$ spectrum and the  $yim$ fit  to the $ntr$ spectrum.}
  \label{fig:blackholefit}
\end{figure}
\begin{figure}
\centering
\includegraphics[scale=1.0]{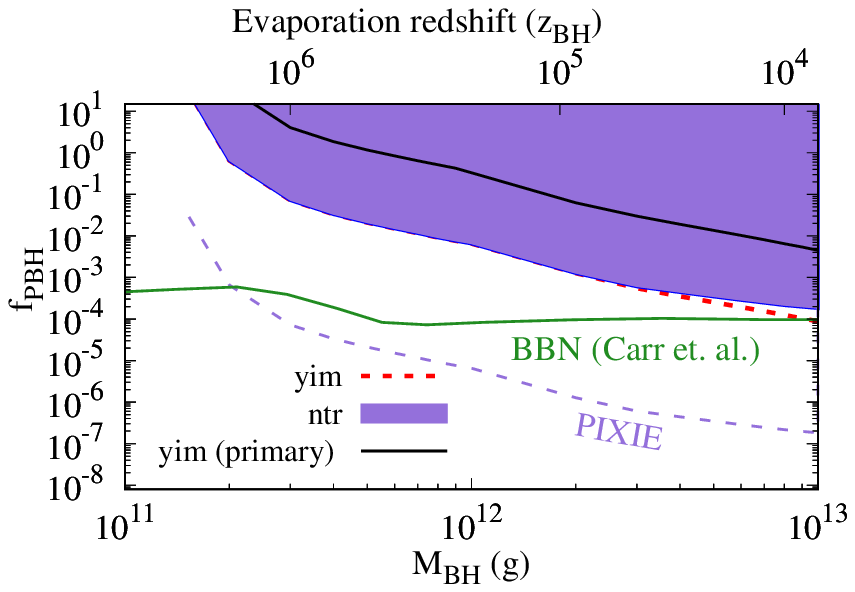}
\caption{Constraints on the initial energy density in  black holes (as a
  fraction $f_{\rm PBH}$ of total stable cold dark matter) as a function of
  mass. The projection for PIXIE-like \cite{Pixie2011} experiment with constraints stronger by a factor of 1000 is also shown. The BBN constraints are taken from Fig. 9 of \cite{LSHLC2019}, which is adapted from \cite{CKSY2010}. The evaporation redshift ($z_{\rm BH}$) of the black hole is
  defined as the redshift at which the mass of the black hole has reduced
  by a factor of e (=2.718). We assume monochromatic black hole mass
  function. The shaded region is excluded by COBE-FIRAS data.}
\label{fig:blackholeconst}
\end{figure}
\par
In Fig. \ref{fig:blackholefraction}, the fraction of the mass of the black
hole yet to be evaporated and the corresponding temperature of the black
hole is plotted as a function of redshift for different initial mass. A
black hole of mass $1 0^{13}$g decays around $z\approx 7000$ while
$10^{12}$g mass black hole decays around  $z\approx 2\times 10^5$. In
Fig. \ref{fig:blackholespectrum}, we plot the instantaneous spectra, after
hadronization of quarks/gluons and decay of unstable particles, of emitted
electromagnetic particles for black holes with mass $10^{12}$g and
$10^{13}$g, at  redshift $z=2\times 10^5$. At this redshift the  $10^{13}$g mass black holes are completely intact as $t<t_{\rm BH}$ (or $z>z_{BH}$). Most of the
photons, electrons  and positrons are found around $\sim$100 MeV. This is
due to the pion decay to photons and electrons-positrons after
hadronization. Charged pions first decay to muons which subsequently decay
to electrons and positrons, while neutral pions decay to photon
pairs. Therefore, the average energy of the electrons and positrons is
slightly less compared to the photons. The bump at high energies,
corresponding to  $\sim x_{\rm{bh},s}T_{\rm{BH}}$,  just below the exponential fall-off is due to the primary emission of photons, electrons, and positrons. In the high energy tail, electron and photon spectrum converge as the emission becomes close to thermal.  \par
\hspace{1cm} 
In Fig. \ref{fig:blackholefit}, we show the spectral distortions from
evaporating black holes calculated by 
evolving the high energy particle cascade until all energy is dissipated
and the spectral distortion shape is frozen. For reference we also show the
spectral distortions obtained in the $yim$ approximation. We also plot the spectral distortion
signal with just the primary photon contribution in the $yim$ approximation (used in most of the previous works). 
There is a factor $\approx$ 50 amplification coming by just taking into
account the fact 
  that particles other than the photons are also emitted. From
Eq. \ref{particle fraction}, we can check that taking into account emission
of all standard model particles  (with the exception of top quark, W, Z bosons and
higgs) gives $f(M_{\rm BH})\sim 12$ while the contribution from primary
photons is just 0.12. Nearly half of the energy of emitted quarks and
gluons is converted to secondary electromagnetic particles, with rest of
the energy lost to neutrinos \cite{DM2010}, which explains the factor 50
amplification. The actual non-thermal distortion shape  for $10^{13}$g black hole
has a longer tail due to the high energy photons (being created at $z$
$\lesssim$10000) compared to the  $yim$-approximation  which
makes the amplitude of the signal at the maximum (as well as minimum) of
the distortion comparatively smaller. This can also be seen from the residual signal
after subtracting best fit $yim$ (i.e. trying to approximate the actual
$ntr$ distortion with a $y$+$i$+$\mu$ distortion fit) signal from the $ntr$ signal. The details
of the fitting procedure are described in  \cite{AK2018}. For $10^{12}$g and less massive black holes, the CMB spectral distortions (being created at $z$ $\sim 10^5$) thermalize and converge to $yim$ distortions. 
\par
\hspace{1cm}
In Fig. \ref{fig:blackholeconst}, we plot the spectral distortion
constraints on the fraction of energy density in the primordial black hole
as a function of the mass of the black hole. We define the fraction of primordial black holes w.r.t CDM,
\begin{equation}
 f_{PBH}=\frac{\rho_{{PBH}}(M_{\rm BH})}{\rho_{\rm{cdm}}},
\end{equation} 
  where $\rho_{\rm{cdm}}$ is the stable cold dark matter energy density and
  $\rho_{{PBH}}(M_{\rm BH})$ is the  primordial black hole energy density
  with initial mass $M_{\rm BH}$. To derive the constraints, we have used
  the COBE-FIRAS \cite{F1996} data. We fit the COBE-FIRAS residuals to the CMB
  spectral distortion templates calculated for primordial black holes of
  different mass \cite{AK2019}. Following COBE-FIRAS procedure to derive
  constraints on $y$ and $\mu$ distortions, we also take into account the
  possible deviation of the actual CMB blackbody temperature from 2.725 K
  and  the Galactic foregrounds. For higher mass black holes (decaying at
  smaller redshifts), the ratio of energy density injected into the CMB to
  that of the energy density of the CMB increases as the CMB energy density
  $\propto (1+z)^4$ while $\rho_{\rm BH}$ $\propto (1+z)^3$. Therefore, the
  spectral distortion constraints get stronger for higher mass black
  holes. At redshifts of $z\approx 2\times 10^6$, photon number changing
  processes become effective \cite{ks2012} and constraints become
  exponentially weaker. Constraints from the non-thermal distortion diverge
  from the thermal distortions at higher mass as thermalization is less
  efficient at lower redshifts where the higher mass black holes evaporate. We also show the projections for PIXIE \cite{Pixie2011} and current BBN constraints from \cite{KY2000,CKSY2010}.
\section{Dark matter decay to standard model particles}
\begin{figure}
\centering
\includegraphics[scale=1.0]{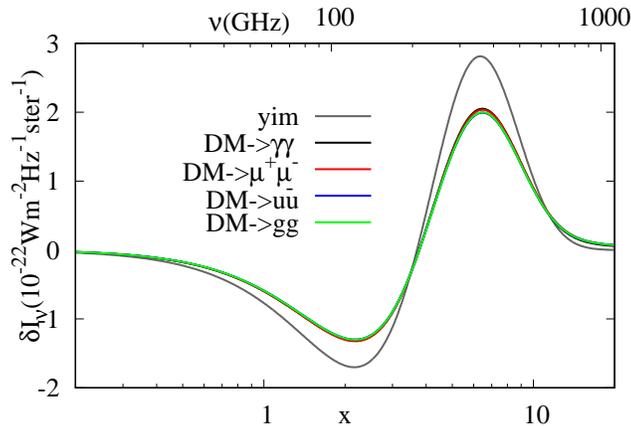}
\caption{Spectral distortions  for 10 GeV dark matter decaying to photon, muon, up quark and gluon pairs. For photon channels $f_X=0.0003$. For other channels, $f_X$ is
  scaled such that total electromagnetic energy injection in these cases is
  equal to that of decay to photons. The decay redshift is $z_X$=20000. For comparison, we show the spectral distortion in the $yim$ approximation assuming all energy goes into heating.}
\label{fig:pmugspectrum}
\end{figure} 

In this section, we expand the results of \cite{AK2019} to additional decay
channels. We consider dark matter decay to neutral pions,
neutral pion-photon pair, muon-antimuon pair, and charged pions for dark
matter mass $300~{\rm MeV} \lesssim m_X \lesssim 2~{\rm  GeV}$. At around   $m_X
\sim  600~{\rm MeV} - 1~{\rm GeV}$, free quarks and gluons are emitted
instead of composite pions. The quarks and gluons after
hadronization produce a spectrum of electrons, positrons, photons,
neutrinos and stable hadrons. We ignore the  energy
 going into stable hadrons (protons/antiprotons,
deuterium/anti-deutrium, Fig 4 of \cite{DM2010}) since interactions of
hadrons with baryon-photon plasma is not included in our code at present. The amount of energy
going into stable hadrons 
depends on the channel and can be a maximum of $\sim 25\%$ of the energy going into
electrons and photons for the pure up, down, and charm quarks and gluon
channels and much smaller in other channels. Our constraints are therefore
slightly conservative for these channels. We plan to include this
in the future. We use PYTHIA results (electromagnetic
spectrum only) \cite{DM2010} for the spectrum of the  standard model
particles from dark matter decay.  We can write the energy density ($E$) injection rate   as,
\begin{equation}
 \frac{dE}{dt}=\frac{f_X \rho_{\rm DM}}{\tau_X} \exp(-t /\tau_X)
\end{equation} 
  where $\tau_X$ is the particle lifetime, $z_X$ is the
  redshift at proper time $t=\tau_X$, $\rho_{\rm DM}(z)=(1+z)^3\rho_{\rm DM}(z=0)$ is the non-decaying
  dark matter energy density at redshift
  $z$ and $f_X$ is the ratio of the initial energy density of decaying dark matter
to that of the non-decaying component. 

\begin{figure}
\centering
\includegraphics[scale=1.0]{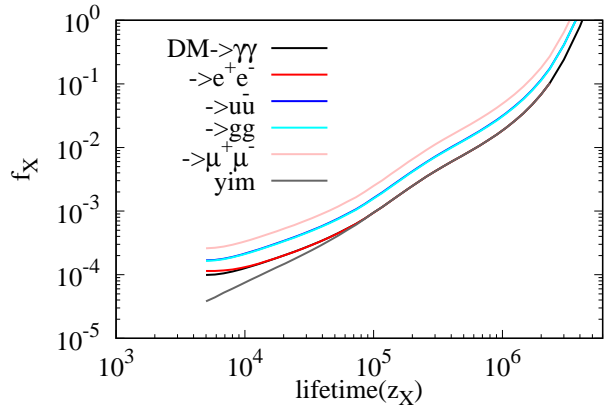}
\caption{Constraints on fraction of decaying dark matter (with
  mass $\sim 1$ GeV and higher), $f_X$, for decay
  to various standard model particles as a function of lifetime. For
  $m_X\gtrsim 1~{\rm GeV}$ the constraints become independent of $m_X$.}
\label{fig:particleconst}
\end{figure}

It was shown in \cite{AK2019} that for dark matter of mass ($m_X$)
$\gtrsim$ 1 GeV decaying to electron-positron pairs or photon pairs,
spectral distortions  are independent of dark matter mass. This
is because for photon energy $\gtrsim$ 1 GeV, the relevant scattering
processes (photon pair production, photon-photon elastic scattering and
inverse Compton scattering) are extremely fast compared to the Hubble
rate. These processes immediately produce a broad, almost universal low
energy photon spectrum below the threshold of these processes, irrespective
of the energy of the injected photon, which makes the spectral distortions and
the constraints derived from spectral distortions independent of the
injected particle spectrum above a redshift dependent energy threshold. This
pattern is also seen in the universal constraints on dark matter decay from
big bang nucleosynthesis \cite{KM1995} and in the energy deposition
fraction from dark matter decay or annihilation around recombination
(Fig. 4 of \cite{Slatyer:2009yq}). We show in Fig. \ref{fig:pmugspectrum},
the spectral distortion for a 10 GeV dark matter decaying to different
standard model particles. The shapes of spectral distortions are almost
identical to that of decay to photons. The amplitude is however different
for different decay channels and is proportional to the energy going into
electromagnetic particles i.e photons, electrons and positrons. In
particular, energy lost to neutrinos depends on the decay channel. Hence,
constraints on dark matter decay to any standard model particle are
sensitive to the decay channel and for dark matter mass $\gtrsim$ few GeV
are  obtained by scaling the constraints from decay to photons with
branching fraction to electromagnetic particles. In
Fig. \ref{fig:particleconst}, we show constraints on the fraction of decaying
dark matter  for muon, up quark and gluon channels. Others quarks have
similar constraints as up quark while tau lepton channel has similar
constraints as the muon channel. Leptonic channels have the  weakest
constraints primarily producing neutrinos while electromagnetic channels
have the strongest constraints. \par
\hspace{1cm}
We also show BBN constraints for dark matter decaying to photon
pairs as a representative case for comparison, extending the
results of  \cite{AK20192} to energies $\gtrsim ~{\rm GeV}$. The strongest
constraints come from photo-dissociation
of helium-4 to  helium-3, resulting in helium-3 overproduction.  The
constraints become stronger with increasing redshifts as CMB photons have
higher energy at higher redshifts. Therefore, the CMB photons can be
boosted to higher energies by electrons and positrons produced in the
cascade. With more number of photons above the threshold of helium-4
destruction, $E_{\rm th}^{\rm He}$, we get stronger constraints. At even higher redshifts,
electron-positron pair production threshold on the CMB photons becomes
comparable to the photo-dissociation threshold of helium and the initial
high energy photons get converted into electron-positron pairs instead of
dissociating helium, since the number density of CMB photons is $\sim 10$ orders of
magnitude larger compared to the helium-4 nuclei. The inverse Compton scattered photons boosted by these
electrons and positrons have energies less than $E_{\rm th}^{\rm He}$. Thus, even though the energy of initial injected photon
maybe larger than  $E_{\rm th}^{\rm He}$, these photons
are immediately degraded to lower energy photons below $E_{\rm th}^{\rm
  He}$ and the constraints become exponentially weaker. We also show PIXIE
\cite{Pixie2011} projection which is assumed to be 1000 times stronger
compared to COBE-FIRAS constraint. Thus, even though at present the BBN
constraints are stronger compared to the CMB spectral distortion
constraints,   in the future the opposite may become true.

\begin{figure}[!tbp]
  \begin{subfigure}[b]{0.4\textwidth}
    \includegraphics[scale=0.8]{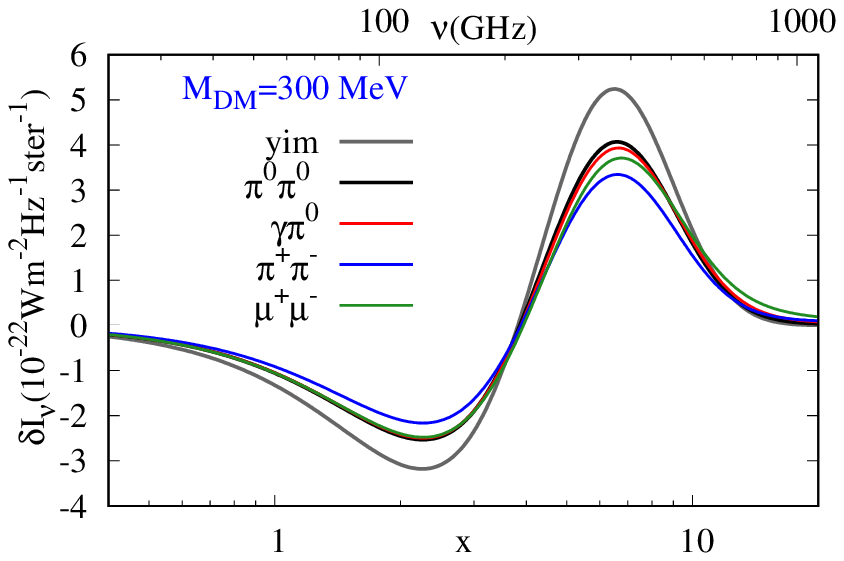}
  \end{subfigure}\hspace{30 pt}
  \begin{subfigure}[b]{0.4\textwidth}
    \includegraphics[scale=0.8]{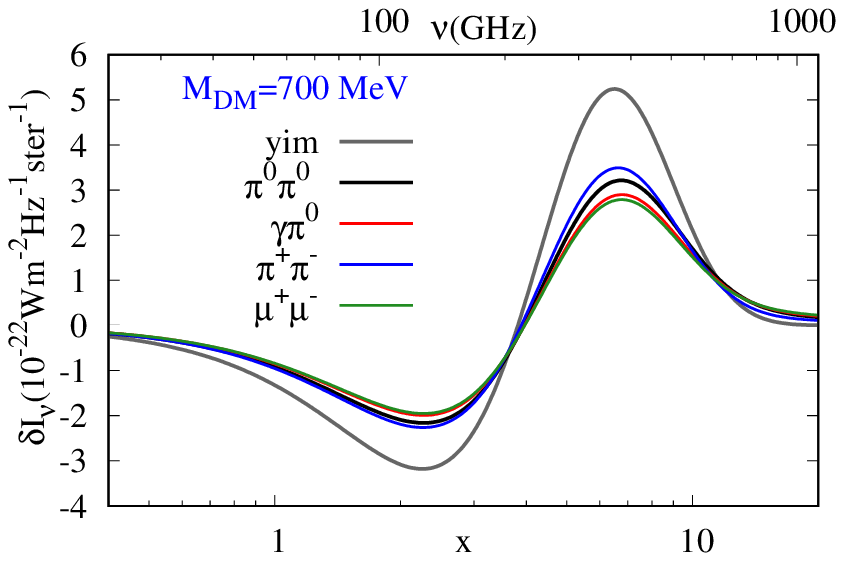}
    \end{subfigure}\\
    
    \begin{subfigure}[b]{0.4\textwidth}
    \includegraphics[scale=0.8]{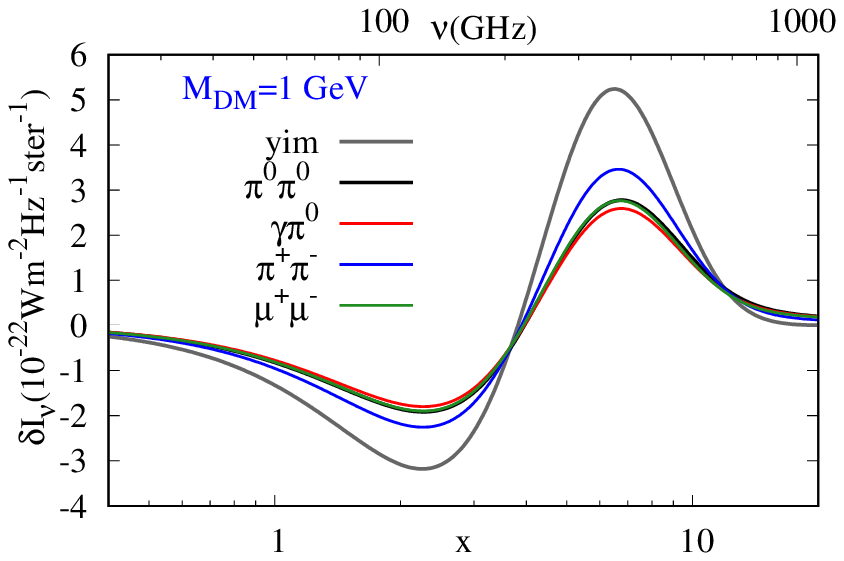}
  \end{subfigure}\hspace{30 pt}
   \caption{Spectrum for dark matter decaying to various channels with
     different dark matter mass,
     $f_X$=0.0003 and $z_X$=10000. The charged pion and muon channel curves
     have been scaled such that the total electromagnetic energy injection
     for each cases is identical. For comparison we show the spectral distortion in the $yim$ approximation assuming all energy goes into heating.}
  \label{fig:spectrumchannels}
\end{figure}

\begin{figure}[!tbp]
  \begin{subfigure}[b]{0.4\textwidth}
    \includegraphics[scale=0.8]{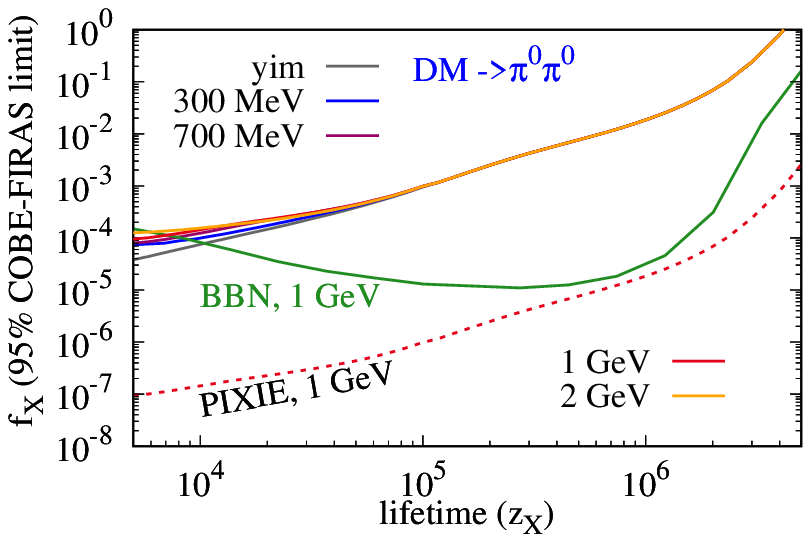}
  \end{subfigure}\hspace{30 pt}
  \begin{subfigure}[b]{0.4\textwidth}
    \includegraphics[scale=0.8]{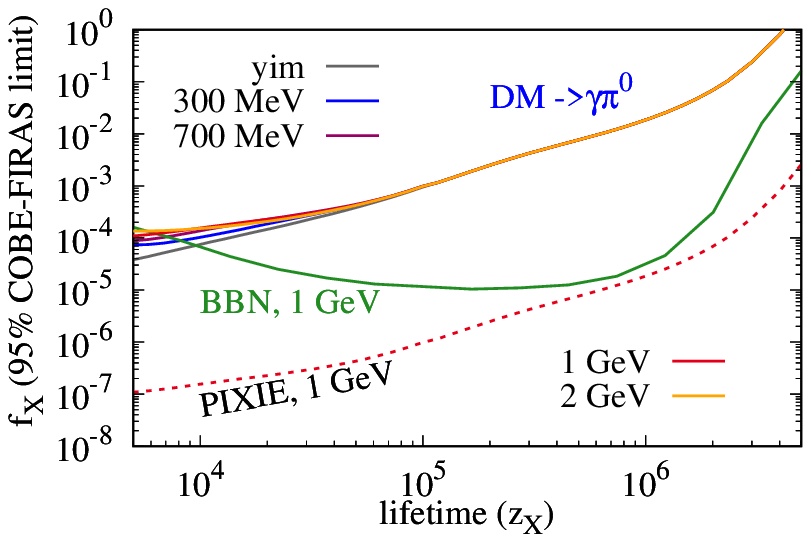}
    \end{subfigure}\\
    
    \begin{subfigure}[b]{0.4\textwidth}
    \includegraphics[scale=0.8]{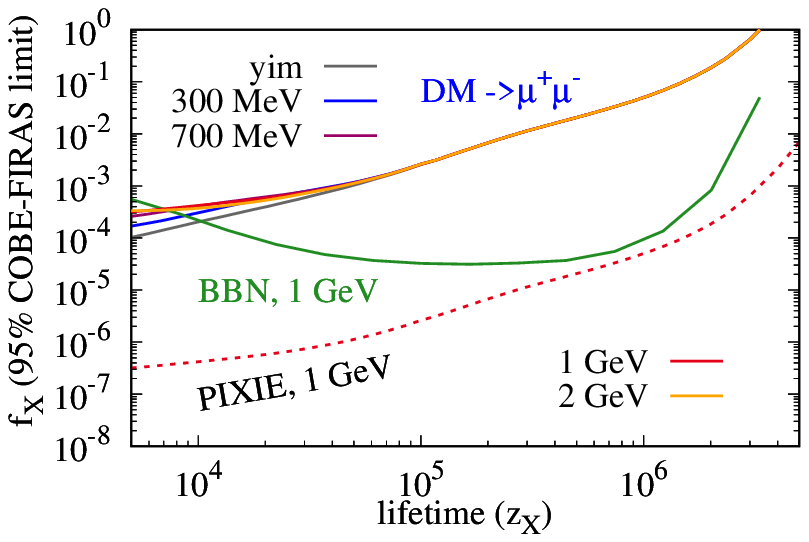}
  \end{subfigure}\hspace{30 pt}
  \begin{subfigure}[b]{0.4\textwidth}
    \includegraphics[scale=0.8]{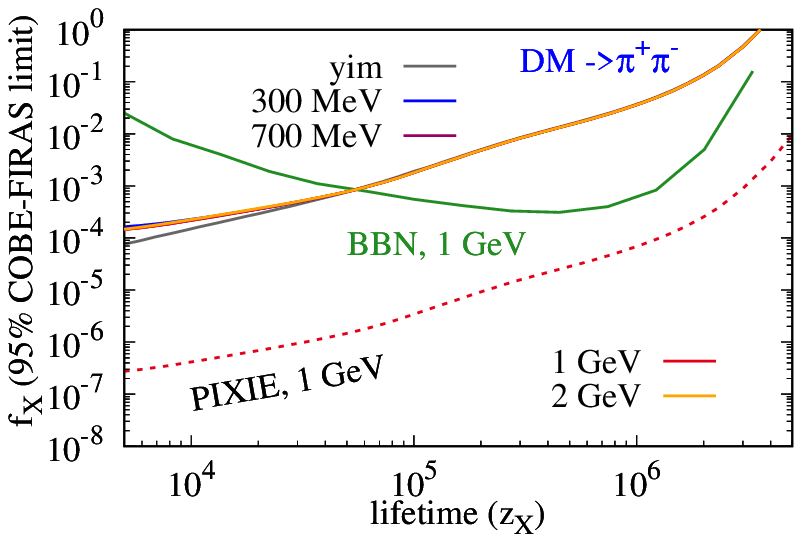}
   \end{subfigure}
   \caption{$95\%$ COBE-FIRAS upper limits on $f_X$ for dark matter decay to different particles for different $m_X$ as
      a function of decay redshift $z_X$.
  Also shown are the constraints assuming all energy going to $yim$ distortions. We also show BBN constraints and PIXIE projection for a representative case of dark matter decay with mass 1 GeV.}
  \label{fig:constchannels}
\end{figure}

\begin{figure}[!tbp]
  \begin{subfigure}[b]{0.4\textwidth}
    \includegraphics[scale=0.8]{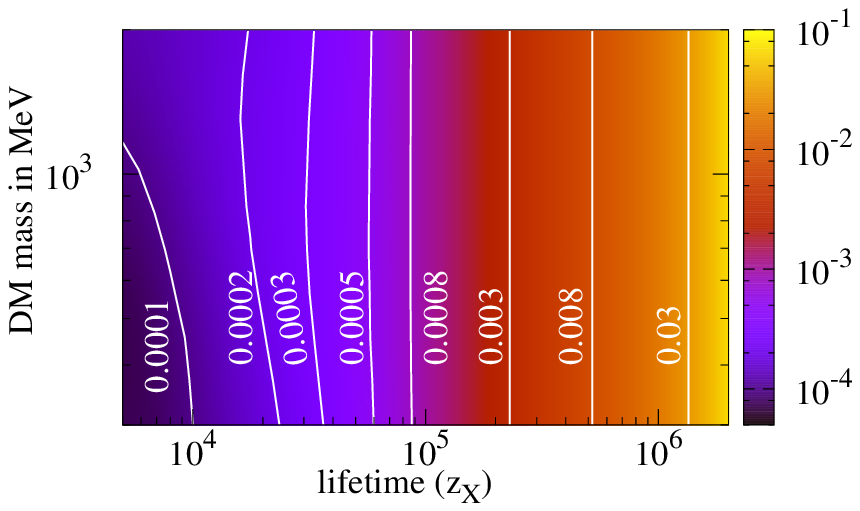}
  \end{subfigure}\hspace{30 pt}
  \begin{subfigure}[b]{0.4\textwidth}
    \includegraphics[scale=0.8]{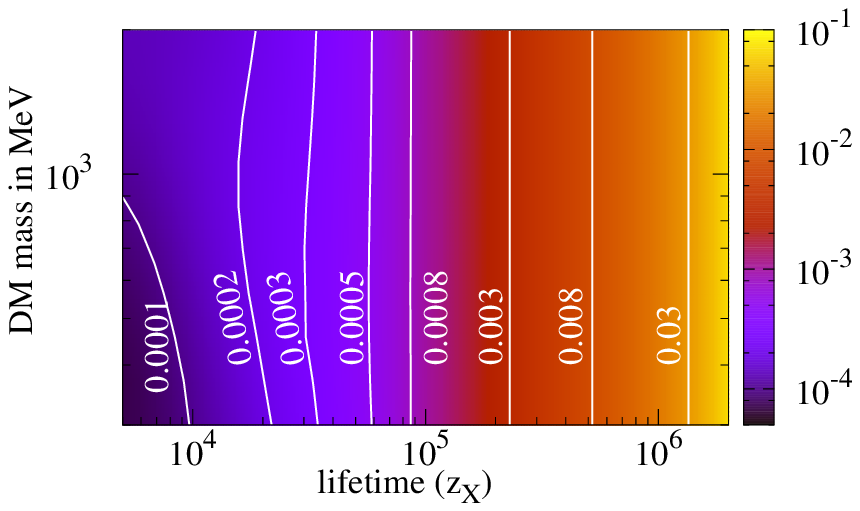}
    \end{subfigure}\\
    
    \begin{subfigure}[b]{0.4\textwidth}
    \includegraphics[scale=0.8]{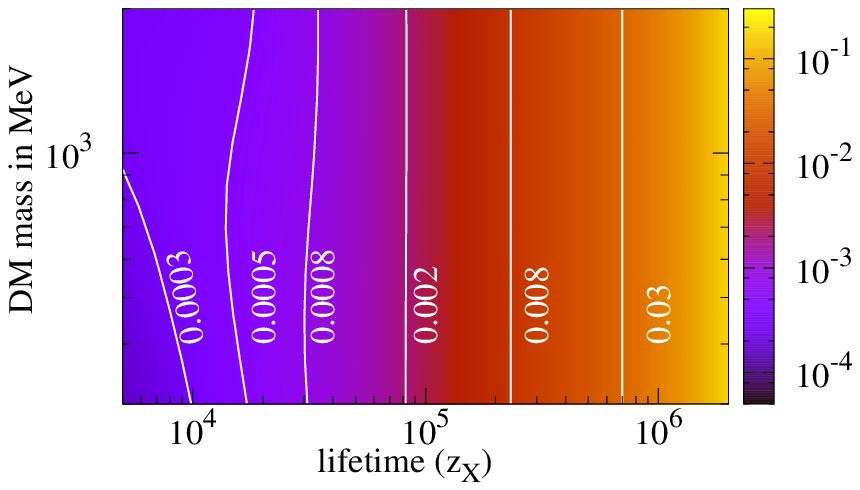}
  \end{subfigure}\hspace{30 pt}
  \begin{subfigure}[b]{0.4\textwidth}
    \includegraphics[scale=0.8]{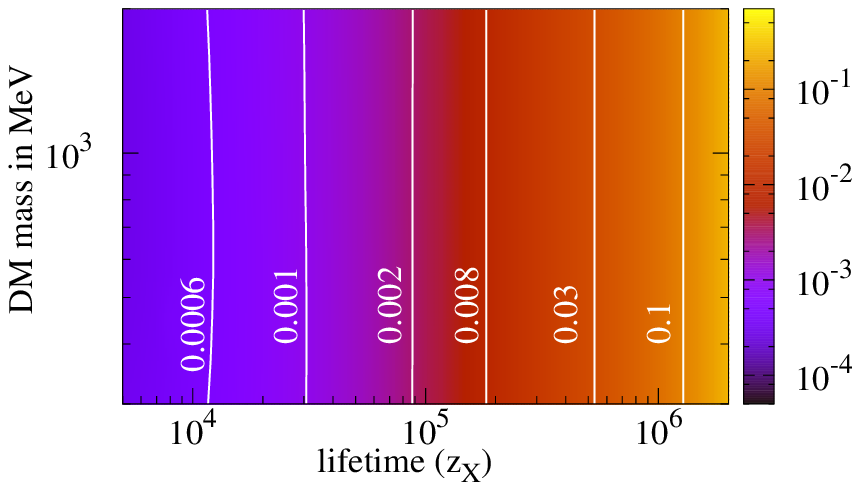}
   \end{subfigure}
   \caption{$95\%$ COBE-FIRAS upper limits on $f_X$ for dark matter decay to neutral pions (upper left), photon-neutral pion pairs (upper right), muons (lower left), charged pion pairs (lower right) in the $z_X-m_X$ plane.}
  \label{fig:constchannels1}
  \end{figure}
  For  $m_X\lesssim 2~{\rm GeV}$, we consider dark
  matter decay to neutral pions, photon-neutral pion, charged pions, and
  muons. These channels have been considered for the detection of dark
  matter through pair annihilation to gamma rays \cite{GPR2017,BGW2017}. We
  use the electron, positron, and photon spectrum provided in the
  references \cite{GPR2017,BGW2017}. In Fig. \ref{fig:spectrumchannels}, we
  show the CMB spectral distortions for the  pion and muon
  decay channels for $z_X$=10000. We can see that the shape of the distortion depends upon the decay
  channel since the  initial injected photon, electron-positron pairs spectra are
  different for different decay channels. In Fig. \ref{fig:constchannels}
  and \ref{fig:constchannels1}, we give the constraints on dark matter
  decaying to neutral pions, photon-neutral pion pairs, muons, charged
  pions for different values of dark matter mass. For the neutral pion pair and
  photon-neutral pion pair channels, all of the mass-energy of the dark
  matter is released in the form of
  electromagnetic energy since  the neutral pion decays to two photons. For the muon
  and charged pion channels a significant fraction of energy is lost to
  neutrinos. The fraction of deposited electromagnetic energy  is $\sim$35
  percent for the muon channel and $\sim$ 25 percent for the charged pions, relaxing the constraints compared to the other channels. 
  \section{Decay of cosmic strings}
  Cosmic strings are one dimensional topological defects produced in the
  early universe  during the cosmological phase transitions that break U(1) symmetry spontaneously \cite{V1985,VS1994}. The cosmic strings can be superconducting, carrying currents, and can radiate bursts of electromagnetic energy. The amount of energy released and the energy spectrum of particles produced in the electromagnetic bursts is a function of string tension and current \cite{W1985,OTW1986,VV1987,CSV2012,CSSDV2012}. The power emitted in electromagntic energy is \cite{VV1987} $P_{EM}=\Gamma_{EW}I\sqrt{\mu}$, where $I$ is the current in the  string, and $\mu$ is the string tension. They can also radiate gravitational waves \cite{VV1985} with power  $P_{GW}=\Gamma_{GW}G\mu^2$, where $\Gamma_{GW}$ and $\Gamma_{EM}$ (in case of electromagnetic emission) depend on the shapes of string loops \cite{VV1985,VV1987}.  The radiated electromagnetic spectrum has a cutoff, which comes from demanding that the energy emitted does not backreact on cosmic string dynamics. The cutoff is given by \cite{VV1987,MN2013}, $\omega_c=2\pi f_c=\mu^{3/2}I^{-3}l^{-1}$. The  spectrum of the electromagnetic radiation is given by, $\frac{dE}{df}\propto f^{-2/3}$. The crossover between electromagnetic and gravitational dominated energy loss is given by a curve in the $G\mu-I$ plane. For $I>I^{*}$, where $I^{*}=\frac{\Gamma_{GW}G \mu^{3/2}}{\Gamma_{EW}}$, the energy loss is mainly by electromagnetic radiation. We consider $\Gamma_{EW}\sim 10, \Gamma_{GW}\sim 50$ in this paper following \cite{MN2013}. These numbers are derived for string trajectories found in \cite{KT1982} for non self-intersecting strings. \par
  \hspace{1cm}
  We consider decay of cosmic strings with short-lived loops following \cite{MN2013} as a matter of simplicity. The rate of shrinkage is given by,
  \begin{equation}
  \frac{dl}{dt}=-\Gamma_{eff},
  \end{equation}
   where $\Gamma_{eff}=\Gamma_{GW}G\mu+\Gamma_{EM}\frac{I}{\sqrt{\mu}}$.
  Assuming the length of loop at formation time $t_i$ to be $\alpha t_i$, the length of loop decreases as,
  \begin{equation}
  l(t)=\alpha t_i-\Gamma_{eff}(t-t_i)
  \end{equation}
  The lifetime of a loop is given by, $\tau=\frac{\alpha}{\Gamma_{eff}}t_i$. For a short-lived loop, $\alpha=\Gamma_{eff}$. The formation rate of loops at time t is given by \cite{MN2013}, $\frac{dn}{dt}\sim\frac{1}{\gamma^2\alpha t^4}$ while number density of decaying loops for small loops is given by $n_d(t)\sim \frac{1}{\gamma^2\alpha t^3}$ with $\gamma=0.27$ in the radiation dominated era. The electromagnetic energy density injected per unit time is given by,
  \begin{equation}
  \frac{dQ}{dt}=P_{EM}n_d(t)
  \end{equation}
  The spectral distortion constraints obtained in \cite{MN2013} assume the
  distortions to be of thermal nature regardless of the spectrum of
  injected energy. In this work, we will use the information of full
  injected spectrum to calculate the distortion spectrum. The photons
  emitted by strings can have higher or lower energy compared to the CMB
  photons. For a high current, a string can radiate a lot of soft
  photons. Therefore, we have to take into account photon injection in the
  form of low energy photons (lower than the average CMB photon
  energy). For photon injection with 0.1$\lesssim x\lesssim $ 10, where
  $x=(\frac{E_{\gamma}}{T_{\rm{CMB}}})$, Compton scattering is the dominant
  process \cite{ks2012,C2015}. For x$<$0.1, photon non-conserving processes
  Bremsstrahlung (BR) and double Compton scattering (DC) are important
  \cite{Chluba:2011hw,ks2012,C2015}. The soft photons are absorbed through
  these processes, heating the electrons. Therefore, for these low energy
  photon injections, we will get $yim$ distortions, while for photon with
  energy 0.1$\lesssim x \lesssim $10, distortion shape is much richer
  \cite{C2015}. The transition from Compton scattering dominated regime to
  BR, DC dominated region can be seen in Fig. 3 of \cite{C2015} and is in
  reality not sharp but gradual. In our calculations, we assume a sharp
  crossover between pure photon non-conserving processes (BR and DC) and
  Compton scattering at x=0.01 to simplify calculations. This is a good
  approximation as at $x\lesssim 0.01$, the BR absorption rate is faster
  compared to the Hubble rate (Fig. 3 of \cite{ks2012}), implying almost
  complete absorption of these photons. In our approximate treatment,
  photons at x$>0.01$ are evolved with Compton scattering which after
  crossing x=0.01 boundary are converted to heat. This is a good
  approximation for COBE-FIRAS data, where we observe CMB spectrum only at
  $x>1.2$. We have checked that the spectral distortion shapes for $x>1$ are not sensitive to the choice of this boundary.    \par
  \hspace{1cm}
  \begin{figure}[!tbp]
  \begin{subfigure}[b]{0.4\textwidth}
    \includegraphics[scale=0.8]{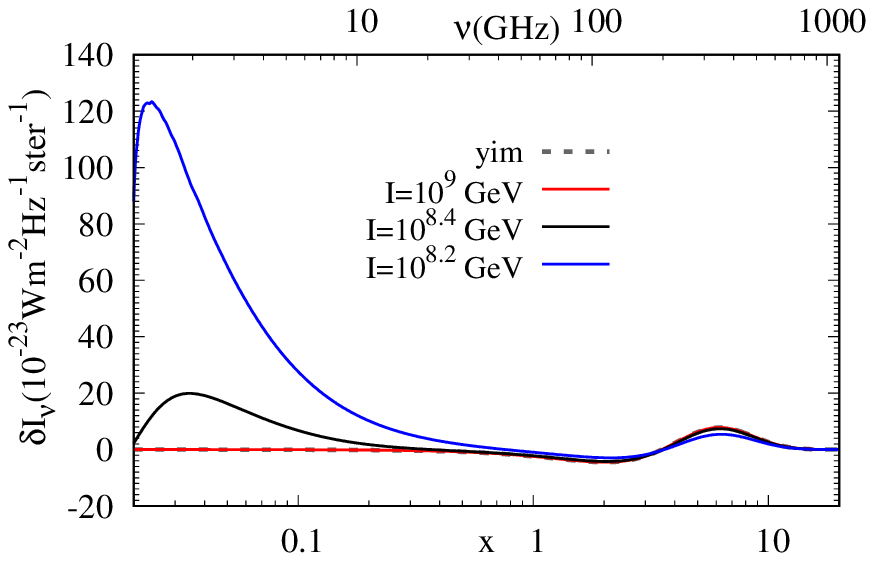}
    \caption{}
    \label{fig:cobe}
  \end{subfigure}\hspace{30 pt}
  \begin{subfigure}[b]{0.4\textwidth}
    \includegraphics[scale=0.8]{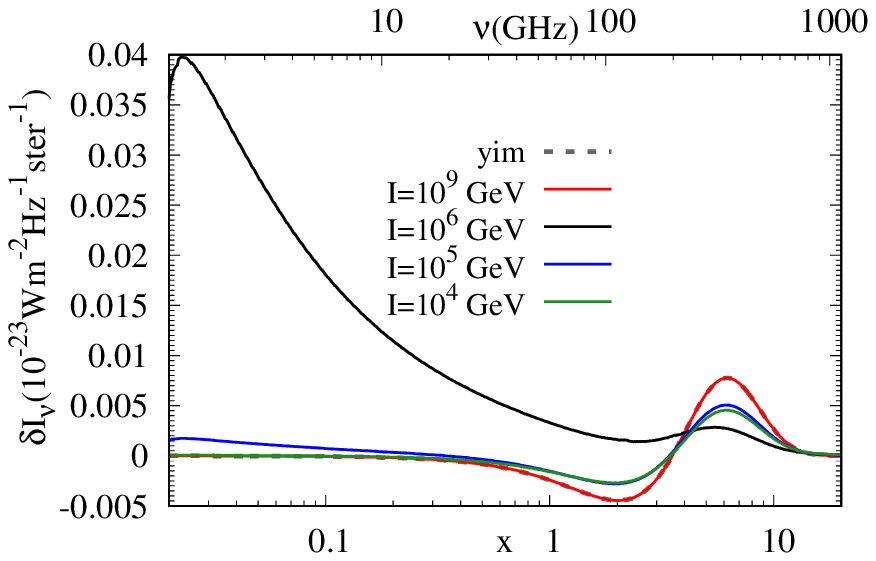}
   \caption{}
   \label{fig:pixie}
    \end{subfigure}\\
    
   \caption{Comparison of spectral distortion shape with $yim$ approximation with constant dimensionless string tension (a) $G\mu=10^{-8.6}$, (b) $G\mu=10^{-11.6}$ and varying currents.}
  \label{fig:strings}
\end{figure} 
\begin{figure}
\centering
\includegraphics[scale=1.0]{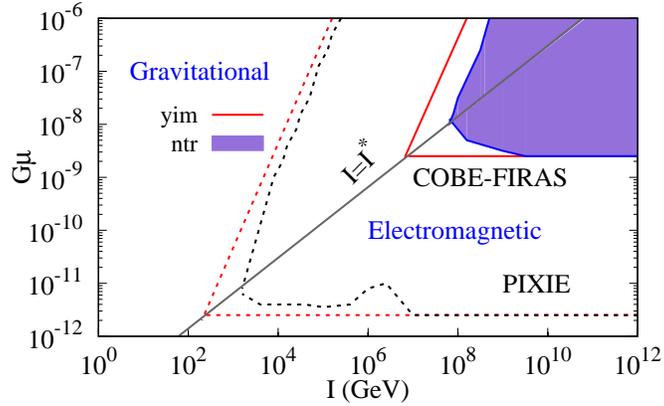}
\caption{Exclusion plot for dimensionless string tension vs current for
  COBE-FIRAS \cite{F1996} (black shaded region) and projections for PIXIE
  \cite{Pixie2011} (dashed lines). For the string parameters corresponding to the bump in PIXIE projections at $I\approx 10^6$ GeV, photons are emitted in the energy range with $0.01<x<1$. In this energy range, bremsstrahlung process is inefficient and photons have high chance to survive rather being absorbed and converted to heat, making constraints significantly weaker, since most of these photons are out of the CMB (COBE-FIRAS) band. }
\label{fig:exclusion}
\end{figure} 
In Fig. \ref{fig:strings}, we compare the actual spectral distortion shape
and $yim$  shape for photon injection from EM-dominated string decay with
constant tension and different currents. For $I \gtrsim 10^9$ GeV, the
emitted soft photons ($x<0.01$) heat up the electrons immediately. For
slightly smaller currents, photons are produced in the range $0.01<x<10$,
which have higher chance to survive while giving a small fraction of energy
to heat by absorption and through Compton scattering. This relaxes the
COBE-FIRAS \cite{F1996} constraints on allowed string tension as shown in
Fig. \ref{fig:cobe} and \ref{fig:exclusion} and produces the big bump at
$I=10^6$GeV  in the PIXIE projections. For still lower current, photons
with relativistic energies are produced (Fig. \ref{fig:pixie}) which give
distortions with lower amplitude and significantly more  energy in the  high energy
tail compared to the $yim$ distortions. In Fig. \ref{fig:exclusion}, we
give the spectral distortion constraints and projections for PIXIE
experiment assuming a factor of 1000 improvement in sensitivity for both
EM- and GW-dominated string decay over COBE-FIRAS. For reference, we also
show the constraints in the $yim$ approximation i.e. assuming all radiated
energy goes into heat. We see that the actual constraints are significantly
relaxed and $yim$ approximation can be off by almost  an order of
magnitude. For a given value of current, $I$, as we increase
the string tension, the fraction of decay  energy going into gravitational
waves (photons) increases (decreases) and the CMB spectral distortions
constraints become weaker. The high tension part of the parameter space is
 strongly constrained by the limits on the stochastic gravitational
wave background 
region from the pulsar timing arrays \cite{MN2013}.

\section{Conclusions}
In this work, we have studied the spectral distortion constraints from
evaporation of primordial black holes, dark matter decay to standard model
particles and decay of cosmic strings. We solve the high energy particle
cascade from black hole evaporation, taking into account secondary particles
produced from hadronization of emitted primary particles, and obtained
constraints which are stronger by almost a factor of 50 compared to the
previous works which ignored the secondary particles. Primordial black
holes of mass $\sim 10^{13}$g (decaying at $z\lesssim 10^4$) inject
relativistic particles in the  background baryon-photon plasma, which
results in 
relativistic non-thermal spectral distortions having higher intensity in the  high
energy tail compared to the non-relativistic thermal ($yim$) spectral
distortions. The lower amplitude of spectral intensity relaxes the
constraints compared to the $yim$ approximation by a factor of 2. For black
hole of mass $10^{12}$g and below, the spectral distortions thermalize to
$\mu$-type distortions.\par
\hspace{1cm}
  We explicitly show that for decaying dark matter with mass of the order
  of 10 GeV and above, the spectral distortion shape is independent of dark
  matter mass and the constraints for decay to various standard model
  particle channels is simply a function of branching fraction to
  electromagnetic particles. For sub-GeV mass dark matter, the spectral
  distortions are non-universal having a characteristic shape that is
  sensitive to the decay channel as well as the mass of the dark matter
  particle, and can deviate from $yim$ approximation  by a factor of 2 to 3 in amplitude in the COBE-FIRAS frequency band. \par 
\hspace{1cm}
We also provide constraints on superconducting cosmic strings in the string
tension and current plane, taking into account the emitted photon spectrum
which had been ignored in previous studies. We show that the photon emission
with relativistic energy and in the CMB band has palpable consequences for
both COBE-FIRAS and PIXIE spectral distortion constraints. Taking the
actual shape into account 
relaxes the constraints by almost an order of magnitude in the region of
the parameter space where photons with energy comparable or larger than the
average CMB photon energy are being emitted. The shape of the spectral
distortion is sensitive to the tension and current of the cosmic
string.

The spectral distortions in the $y$-type and $i$-type era ($z\lesssim
10^5$) carry rich information about the potential new physics beyond the
standard model of particle physics. This information is hidden in the
characteristic shapes of the CMB spectral distortions and can, with
precision measurements of the CMB spectrum, distinguish the particular new
physics responsible for creating the distortions. The CMB spectral
distortions, if detected (other than the Sunyaev-Zeldovich effect which has
already been detected), can thus be
used to measure the parameters of the new physics responsible for them,
such as the branching
fractions and mass of the decaying dark matter, mass function of primordial black holes,
and tension and current of the cosmic strings.

\section{Acknowledgements}
 This work was supported by Max Planck Partner Group for cosmology of Max Planck
 Institute for Astrophysics Garching at 
 Tata Institute of Fundamental Research funded by
 Max-Planck-Gesellschaft. This work was also supported by 
Science and Engineering Research Board (SERB) of Department of Science and
Technology, Government of India grant no. ECR/2015/000078. We acknowledge support of the Department of Atomic Energy, Government of India, under project no. 12-R\&D-TFR-5.02-0200.

\bibliographystyle{unsrtads}
\bibliography{spec-cs1} 

\begin{thebibliography}{10}

\bibitem{ASS1998}
J.~A. {Adams}, S.~{Sarkar}, and D.~W. {Sciama}.
\newblock {Cosmic microwave background anisotropy in the decaying neutrino
  cosmology}.
\newblock {\em \mnras}, 301:210--214, November 1998.
\newblock \href {http://arxiv.org/abs/astro-ph/9805108}
  {\path{arXiv:astro-ph/9805108}}, \href
  {http://dx.doi.org/10.1046/j.1365-8711.1998.02017.x} {\path{[DOI]}},
  {\small[\href{http://adsabs.harvard.edu/abs/1998MNRAS.301..210A}{ADS}]}.

\bibitem{Chen:2003gz}
X.~{Chen} and M.~{Kamionkowski}.
\newblock {Particle decays during the cosmic dark ages}.
\newblock {\em \prd}, 70(4):043502, August 2004.
\newblock \href {http://arxiv.org/abs/astro-ph/0310473}
  {\path{arXiv:astro-ph/0310473}}, \href
  {http://dx.doi.org/10.1103/PhysRevD.70.043502} {\path{[DOI]}},
  {\small[\href{http://adsabs.harvard.edu/abs/2004PhRvD..70d3502C}{ADS}]}.

\bibitem{Galli:2013dna}
S.~{Galli}, T.~R. {Slatyer}, M.~{Valdes}, and F.~{Iocco}.
\newblock {Systematic uncertainties in constraining dark matter annihilation
  from the cosmic microwave background}.
\newblock {\em \prd}, 88(6):063502, September 2013.
\newblock \href {http://arxiv.org/abs/1306.0563} {\path{arXiv:1306.0563}},
  \href {http://dx.doi.org/10.1103/PhysRevD.88.063502} {\path{[DOI]}},
  {\small[\href{http://adsabs.harvard.edu/abs/2013PhRvD..88f3502G}{ADS}]}.

\bibitem{AK20192}
Sandeep~Kumar {Acharya} and Rishi {Khatri}.
\newblock {CMB anisotropy and BBN constraints on pre-recombination decay of
  dark matter to visible particles}.
\newblock {\em \jcap}, 2019(12):046, Dec 2019.
\newblock \href {http://arxiv.org/abs/1910.06272} {\path{arXiv:1910.06272}},
  \href {http://dx.doi.org/10.1088/1475-7516/2019/12/046} {\path{[DOI]}},
  {\small[\href{https://ui.adsabs.harvard.edu/abs/2019JCAP...12..046A}{ADS}]}.

\bibitem{Pl2018}
N.~Aghanim et~al.
\newblock {Planck 2018 results. VI. Cosmological parameters}.
\newblock {\em ArXiv e-prints}, July 2018.
\newblock \href {http://arxiv.org/abs/1807.06209} {\path{arXiv:1807.06209}},
  {\small[\href{http://adsabs.harvard.edu/abs/2018arXiv180706209P}{ADS}]}.

\bibitem{SPT2016}
{de Haan} et. al.
\newblock {Cosmological Constraints from Galaxy Clusters in the 2500
  Square-degree SPT-SZ Survey}.
\newblock {\em \apj}, 832(1):95, Nov 2016.
\newblock \href {http://arxiv.org/abs/1603.06522} {\path{arXiv:1603.06522}},
  \href {http://dx.doi.org/10.3847/0004-637X/832/1/95} {\path{[DOI]}},
  {\small[\href{https://ui.adsabs.harvard.edu/abs/2016ApJ...832...95D}{ADS}]}.

\bibitem{ACT2017}
{Louis} et. al.
\newblock {The Atacama Cosmology Telescope: two-season ACTPol spectra and
  parameters}.
\newblock {\em \jcap}, 2017(6):031, Jun 2017.
\newblock \href {http://arxiv.org/abs/1610.02360} {\path{arXiv:1610.02360}},
  \href {http://dx.doi.org/10.1088/1475-7516/2017/06/031} {\path{[DOI]}},
  {\small[\href{https://ui.adsabs.harvard.edu/abs/2017JCAP...06..031L}{ADS}]}.

\bibitem{Hazumi2019}
M.~Hazumi et~al.
\newblock {LiteBIRD: A Satellite for the Studies of B-Mode Polarization and
  Inflation from Cosmic Background Radiation Detection}.
\newblock {\em J. Low. Temp. Phys.}, 194(5-6):443--452, 2019.
\newblock \href {http://dx.doi.org/10.1007/s10909-019-02150-5} {\path{[DOI]}}.

\bibitem{Pixie2011}
A.~{Kogut}, D.~J. {Fixsen}, D.~T. {Chuss}, J.~{Dotson}, E.~{Dwek},
  M.~{Halpern}, G.~F. {Hinshaw}, S.~M. {Meyer}, S.~H. {Moseley}, M.~D.
  {Seiffert}, D.~N. {Spergel}, and E.~J. {Wollack}.
\newblock {The Primordial Inflation Explorer (PIXIE): a nulling polarimeter for
  cosmic microwave background observations}.
\newblock {\em \jcap}, 7:025, July 2011.
\newblock \href {http://arxiv.org/abs/1105.2044} {\path{arXiv:1105.2044}},
  \href {http://dx.doi.org/10.1088/1475-7516/2011/07/025} {\path{[DOI]}},
  {\small[\href{https://ui.adsabs.harvard.edu/abs/2011JCAP...07..025K}{ADS}]}.

\bibitem{CORE2018}
{Di Valentino} et~al.
\newblock {Exploring cosmic origins with CORE: Cosmological parameters}.
\newblock {\em \jcap}, 2018(4):017, Apr 2018.
\newblock \href {http://arxiv.org/abs/1612.00021} {\path{arXiv:1612.00021}},
  \href {http://dx.doi.org/10.1088/1475-7516/2018/04/017} {\path{[DOI]}},
  {\small[\href{https://ui.adsabs.harvard.edu/abs/2018JCAP...04..017D}{ADS}]}.

\bibitem{PICO2019}
{Hanany} et. al.
\newblock {PICO: Probe of Inflation and Cosmic Origins}.
\newblock {\em arXiv e-prints}, page arXiv:1902.10541, Feb 2019.
\newblock \href {http://arxiv.org/abs/1902.10541} {\path{arXiv:1902.10541}},
  {\small[\href{https://ui.adsabs.harvard.edu/abs/2019arXiv190210541H}{ADS}]}.

\bibitem{CMBS42019}
{Carlstrom} et. al.
\newblock {CMB-S4}.
\newblock volume~51, page 209, Sep 2019.
\newblock \href {http://arxiv.org/abs/1908.01062} {\path{arXiv:1908.01062}},
  {\small[\href{https://ui.adsabs.harvard.edu/abs/2019BAAS...51g.209C}{ADS}]}.

\bibitem{PS2015}
Vivian {Poulin} and Pasquale~Dario {Serpico}.
\newblock {Nonuniversal BBN bounds on electromagnetically decaying particles}.
\newblock {\em \prd}, 91(10):103007, May 2015.
\newblock \href {http://arxiv.org/abs/1503.04852} {\path{arXiv:1503.04852}},
  \href {http://dx.doi.org/10.1103/PhysRevD.91.103007} {\path{[DOI]}},
  {\small[\href{https://ui.adsabs.harvard.edu/abs/2015PhRvD..91j3007P}{ADS}]}.

\bibitem{HSW2018}
Marco {Hufnagel}, Kai {Schmidt-Hoberg}, and Sebastian {Wild}.
\newblock {BBN constraints on MeV-scale dark sectors. Part II: Electromagnetic
  decays}.
\newblock {\em Journal of Cosmology and Astro-Particle Physics}, 2018(11):032,
  Nov 2018.
\newblock \href {http://arxiv.org/abs/1808.09324} {\path{arXiv:1808.09324}},
  \href {http://dx.doi.org/10.1088/1475-7516/2018/11/032} {\path{[DOI]}},
  {\small[\href{https://ui.adsabs.harvard.edu/abs/2018JCAP...11..032H}{ADS}]}.

\bibitem{FMW2019}
Lindsay {Forestell}, David~E. {Morrissey}, and Graham {White}.
\newblock {Limits from BBN on light electromagnetic decays}.
\newblock {\em Journal of High Energy Physics}, 2019(1):74, Jan 2019.
\newblock \href {http://arxiv.org/abs/1809.01179} {\path{arXiv:1809.01179}},
  \href {http://dx.doi.org/10.1007/JHEP01(2019)074} {\path{[DOI]}},
  {\small[\href{https://ui.adsabs.harvard.edu/abs/2019JHEP...01..074F}{ADS}]}.

\bibitem{ENS1985}
J.~{Ellis}, D.~V. {Nanopoulos}, and S.~{Sarkar}.
\newblock {The cosmology of decaying gravitinos}.
\newblock {\em Nuclear Physics B}, 259:175--188, September 1985.
\newblock \href {http://dx.doi.org/10.1016/0550-3213(85)90306-2}
  {\path{[DOI]}},
  {\small[\href{http://cdsads.u-strasbg.fr/abs/1985NuPhB.259..175E}{ADS}]}.

\bibitem{EGLNS1992}
J.~{Ellis}, G.~B. {Gelmini}, J.~L. {Lopez}, D.~V. {Nanopoulos}, and
  S.~{Sarkar}.
\newblock {Astrophysical constraints on massive unstable neutral relic
  particles}.
\newblock {\em Nuclear Physics B}, 373:399--437, April 1992.
\newblock \href {http://dx.doi.org/10.1016/0550-3213(92)90438-H}
  {\path{[DOI]}},
  {\small[\href{https://ui.adsabs.harvard.edu/abs/1992NuPhB.373..399E}{ADS}]}.

\bibitem{KM1995}
M.~{Kawasaki} and T.~{Moroi}.
\newblock {Electromagnetic Cascade in the Early Universe and Its Application to
  the Big Bang Nucleosynthesis}.
\newblock {\em \apj}, 452:506, Oct 1995.
\newblock \href {http://arxiv.org/abs/astro-ph/9412055}
  {\path{arXiv:astro-ph/9412055}}, \href {http://dx.doi.org/10.1086/176324}
  {\path{[DOI]}},
  {\small[\href{https://ui.adsabs.harvard.edu/abs/1995ApJ...452..506K}{ADS}]}.

\bibitem{KKMT2018}
Masahiro {Kawasaki}, Kazunori {Kohri}, Takeo {Moroi}, and Yoshitaro {Takaesu}.
\newblock {Revisiting big-bang nucleosynthesis constraints on long-lived
  decaying particles}.
\newblock {\em \prd}, 97(2):023502, Jan 2018.
\newblock \href {http://arxiv.org/abs/1709.01211} {\path{arXiv:1709.01211}},
  \href {http://dx.doi.org/10.1103/PhysRevD.97.023502} {\path{[DOI]}},
  {\small[\href{https://ui.adsabs.harvard.edu/abs/2018PhRvD..97b3502K}{ADS}]}.

\bibitem{F1996}
D.~J. {Fixsen}, E.~S. {Cheng}, J.~M. {Gales}, J.~C. {Mather}, R.~A. {Shafer},
  and E.~L. {Wright}.
\newblock {The Cosmic Microwave Background Spectrum from the Full COBE FIRAS
  Data Set}.
\newblock {\em \apj}, 473:576, December 1996.
\newblock \href {http://arxiv.org/abs/astro-ph/9605054}
  {\path{arXiv:astro-ph/9605054}}, \href {http://dx.doi.org/10.1086/178173}
  {\path{[DOI]}},
  {\small[\href{https://ui.adsabs.harvard.edu/\#abs/1996ApJ...473..576F}{ADS}]}.

\bibitem{FM2002}
D.~J. {Fixsen} and J.~C. {Mather}.
\newblock {The Spectral Results of the Far-Infrared Absolute Spectrophotometer
  Instrument on COBE}.
\newblock {\em \apj}, 581:817--822, December 2002.
\newblock \href {http://dx.doi.org/10.1086/344402} {\path{[DOI]}},
  {\small[\href{http://adsabs.harvard.edu/abs/2002ApJ...581..817F}{ADS}]}.

\bibitem{Sz19701}
R.~A. {Sunyaev} and Y.~B. {Zeldovich}.
\newblock {The interaction of matter and radiation in the hot model of the
  Universe, II}.
\newblock {\em \apss}, 7:20--30, April 1970.
\newblock \href {http://dx.doi.org/10.1007/BF00653472} {\path{[DOI]}},
  {\small[\href{http://adsabs.harvard.edu/abs/1970Ap%26SS...7...20S}{ADS}]}.

\bibitem{dd1982}
L.~{Danese} and G.~{de Zotti}.
\newblock {Double Compton process and the spectrum of the microwave
  background}.
\newblock {\em \aap}, 107:39--42, 1982.
\newblock
  {\small[\href{http://adsabs.harvard.edu/abs/1982A%26A...107...39D}{ADS}]}.

\bibitem{Chluba:2011hw}
J.~{Chluba} and R.~A. {Sunyaev}.
\newblock {The evolution of CMB spectral distortions in the early Universe}.
\newblock {\em \mnras}, 419:1294--1314, January 2012.
\newblock \href {http://arxiv.org/abs/1109.6552} {\path{arXiv:1109.6552}},
  \href {http://dx.doi.org/10.1111/j.1365-2966.2011.19786.x} {\path{[DOI]}},
  {\small[\href{http://adsabs.harvard.edu/abs/2012MNRAS.419.1294C}{ADS}]}.

\bibitem{ks2012}
R.~{Khatri} and R.~A. {Sunyaev}.
\newblock {Creation of the CMB spectrum: precise analytic solutions for the
  blackbody photosphere}.
\newblock {\em \jcap}, 6:38, 2012.
\newblock \href {http://dx.doi.org/10.1088/1475-7516/2012/06/038}
  {\path{[DOI]}},
  {\small[\href{http://adsabs.harvard.edu/abs/2012JCAP...06..038K}{ADS}]}.

\bibitem{Sz1969}
Y.~B. {Zeldovich} and R.~A. {Sunyaev}.
\newblock {The Interaction of Matter and Radiation in a Hot-Model Universe}.
\newblock {\em \apss}, 4:301--316, July 1969.
\newblock \href {http://dx.doi.org/10.1007/BF00661821} {\path{[DOI]}},
  {\small[\href{http://adsabs.harvard.edu/abs/1969Ap%26SS...4..301Z}{ADS}]}.

\bibitem{Is19752}
A.~F. {Illarionov} and R.~A. {Siuniaev}.
\newblock {Comptonization, the background-radiation spectrum, and the thermal
  history of the universe}.
\newblock {\em \sovast}, 18:691--699, June 1975.
\newblock
  {\small[\href{http://adsabs.harvard.edu/abs/1975SvA....18..691I}{ADS}]}.

\bibitem{Bdd1991}
C.~{Burigana}, L.~{Danese}, and G.~{de Zotti}.
\newblock {Formation and evolution of early distortions of the microwave
  background spectrum - A numerical study}.
\newblock {\em \aap}, 246:49--58, June 1991.
\newblock
  {\small[\href{http://adsabs.harvard.edu/abs/1991A%26A...246...49B}{ADS}]}.

\bibitem{Ks2012b}
R.~{Khatri} and R.~A. {Sunyaev}.
\newblock {Beyond y and {$\mu$}: the shape of the CMB spectral distortions in
  the intermediate epoch, $1.5 {\times} 10^{4}\lesssim$z $ \lesssim 2 {\times}
  10^{5}$}.
\newblock {\em \jcap}, 9:016, September 2012.
\newblock \href {http://arxiv.org/abs/1207.6654} {\path{arXiv:1207.6654}},
  \href {http://dx.doi.org/10.1088/1475-7516/2012/09/016} {\path{[DOI]}},
  {\small[\href{http://adsabs.harvard.edu/abs/2012JCAP...09..016K}{ADS}]}.

\bibitem{Chluba:2013vsa}
J.~{Chluba}.
\newblock {Green's function of the cosmological thermalization problem}.
\newblock {\em \mnras}, 434:352--357, September 2013.
\newblock \href {http://arxiv.org/abs/1304.6120} {\path{arXiv:1304.6120}},
  \href {http://dx.doi.org/10.1093/mnras/stt1025} {\path{[DOI]}},
  {\small[\href{http://adsabs.harvard.edu/abs/2013MNRAS.434..352C}{ADS}]}.

\bibitem{AK2018}
Sandeep~Kumar {Acharya} and Rishi {Khatri}.
\newblock {Rich structure of nonthermal relativistic CMB spectral distortions
  from high energy particle cascades at redshifts $z\lesssim 2 {\times}
  10^{5}$}.
\newblock {\em \prd}, 99(4):043520, Feb 2019.
\newblock \href {http://arxiv.org/abs/1808.02897} {\path{arXiv:1808.02897}},
  \href {http://dx.doi.org/10.1103/PhysRevD.99.043520} {\path{[DOI]}},
  {\small[\href{https://ui.adsabs.harvard.edu/abs/2019PhRvD..99d3520A}{ADS}]}.

\bibitem{AK2019}
Sandeep~Kumar {Acharya} and Rishi {Khatri}.
\newblock {New CMB spectral distortion constraints on decaying dark matter with
  full evolution of electromagnetic cascades before recombination}.
\newblock {\em \prd}, 99(12):123510, Jun 2019.
\newblock \href {http://arxiv.org/abs/1903.04503} {\path{arXiv:1903.04503}},
  \href {http://dx.doi.org/10.1103/PhysRevD.99.123510} {\path{[DOI]}},
  {\small[\href{https://ui.adsabs.harvard.edu/abs/2019PhRvD..99l3510A}{ADS}]}.

\bibitem{BHS2005}
Gianfranco {Bertone}, Dan {Hooper}, and Joseph {Silk}.
\newblock {Particle dark matter: evidence, candidates and constraints}.
\newblock {\em \physrep}, 405(5-6):279--390, Jan 2005.
\newblock \href {http://arxiv.org/abs/hep-ph/0404175}
  {\path{arXiv:hep-ph/0404175}}, \href
  {http://dx.doi.org/10.1016/j.physrep.2004.08.031} {\path{[DOI]}},
  {\small[\href{https://ui.adsabs.harvard.edu/abs/2005PhR...405..279B}{ADS}]}.

\bibitem{F2010}
Jonathan~L. {Feng}.
\newblock {Dark Matter Candidates from Particle Physics and Methods of
  Detection}.
\newblock {\em Annual Review of Astronomy and Astrophysics}, 48:495--545, Sep
  2010.
\newblock \href {http://arxiv.org/abs/1003.0904} {\path{arXiv:1003.0904}},
  \href {http://dx.doi.org/10.1146/annurev-astro-082708-101659} {\path{[DOI]}},
  {\small[\href{https://ui.adsabs.harvard.edu/\#abs/2010ARA&A..48..495F}{ADS}]}.

\bibitem{ROM2008}
Massimo {Ricotti}, Jeremiah~P. {Ostriker}, and Katherine~J. {Mack}.
\newblock {Effect of Primordial Black Holes on the Cosmic Microwave Background
  and Cosmological Parameter Estimates}.
\newblock {\em \apj}, 680:829--845, Jun 2008.
\newblock \href {http://arxiv.org/abs/0709.0524} {\path{arXiv:0709.0524}},
  \href {http://dx.doi.org/10.1086/587831} {\path{[DOI]}},
  {\small[\href{https://ui.adsabs.harvard.edu/\#abs/2008ApJ...680..829R}{ADS}]}.

\bibitem{Am2017}
Y.~{Ali-Ha{\"i}moud} and M.~{Kamionkowski}.
\newblock {Cosmic microwave background limits on accreting primordial black
  holes}.
\newblock {\em \prd}, 95(4):043534, February 2017.
\newblock \href {http://arxiv.org/abs/1612.05644} {\path{arXiv:1612.05644}},
  \href {http://dx.doi.org/10.1103/PhysRevD.95.043534} {\path{[DOI]}},
  {\small[\href{http://adsabs.harvard.edu/abs/2017PhRvD..95d3534A}{ADS}]}.

\bibitem{PSCCK2017}
Vivian {Poulin}, Pasquale~D. {Serpico}, Francesca {Calore}, S{\'e}bastien
  {Clesse}, and Kazunori {Kohri}.
\newblock {CMB bounds on disk-accreting massive primordial black holes}.
\newblock {\em \prd}, 96(8):083524, Oct 2017.
\newblock \href {http://arxiv.org/abs/1707.04206} {\path{arXiv:1707.04206}},
  \href {http://dx.doi.org/10.1103/PhysRevD.96.083524} {\path{[DOI]}},
  {\small[\href{https://ui.adsabs.harvard.edu/abs/2017PhRvD..96h3524P}{ADS}]}.

\bibitem{H1974}
S.~W. {Hawking}.
\newblock {Black hole explosions?}
\newblock {\em \nat}, 248:30--31, March 1974.
\newblock \href {http://dx.doi.org/10.1038/248030a0} {\path{[DOI]}},
  {\small[\href{https://ui.adsabs.harvard.edu/abs/1974Natur.248...30H}{ADS}]}.

\bibitem{H1975}
S.~W. {Hawking}.
\newblock {Particle creation by black holes}.
\newblock {\em Communications in Mathematical Physics}, 43:199--220, August
  1975.
\newblock \href {http://dx.doi.org/10.1007/BF02345020} {\path{[DOI]}},
  {\small[\href{https://ui.adsabs.harvard.edu/abs/1975CMaPh..43..199H}{ADS}]}.

\bibitem{P1976}
D.~N. {Page}.
\newblock {Particle emission rates from a black hole: Massless particles from
  an uncharged, nonrotating hole}.
\newblock {\em \prd}, 13:198--206, January 1976.
\newblock \href {http://dx.doi.org/10.1103/PhysRevD.13.198} {\path{[DOI]}},
  {\small[\href{https://ui.adsabs.harvard.edu/abs/1976PhRvD..13..198P}{ADS}]}.

\bibitem{MW1990}
J.~H. {MacGibbon} and B.~R. {Webber}.
\newblock {Quark- and gluon-jet emission from primordial black holes: The
  instantaneous spectra}.
\newblock {\em \prd}, 41:3052--3079, May 1990.
\newblock \href {http://dx.doi.org/10.1103/PhysRevD.41.3052} {\path{[DOI]}},
  {\small[\href{https://ui.adsabs.harvard.edu/abs/1990PhRvD..41.3052M}{ADS}]}.

\bibitem{MW1991}
J.~H. {MacGibbon}.
\newblock {Quark- and gluon-jet emission from primordial black holes. II. The
  emission over the black-hole lifetime}.
\newblock {\em \prd}, 44:376--392, July 1991.
\newblock \href {http://dx.doi.org/10.1103/PhysRevD.44.376} {\path{[DOI]}},
  {\small[\href{https://ui.adsabs.harvard.edu/abs/1991PhRvD..44..376M}{ADS}]}.

\bibitem{TS2008}
Hiroyuki {Tashiro} and Naoshi {Sugiyama}.
\newblock {Constraints on primordial black holes by distortions of the cosmic
  microwave background}.
\newblock {\em \prd}, 78(2):023004, Jul 2008.
\newblock \href {http://arxiv.org/abs/0801.3172} {\path{arXiv:0801.3172}},
  \href {http://dx.doi.org/10.1103/PhysRevD.78.023004} {\path{[DOI]}},
  {\small[\href{https://ui.adsabs.harvard.edu/abs/2008PhRvD..78b3004T}{ADS}]}.

\bibitem{LSHLC2019}
Matteo {Lucca}, Nils {Sch{\"o}neberg}, Deanna~C. {Hooper}, Julien
  {Lesgourgues}, and Jens {Chluba}.
\newblock {The synergy between CMB spectral distortions and anisotropies}.
\newblock {\em arXiv e-prints}, page arXiv:1910.04619, Oct 2019.
\newblock \href {http://arxiv.org/abs/1910.04619} {\path{arXiv:1910.04619}},
  {\small[\href{https://ui.adsabs.harvard.edu/abs/2019arXiv191004619L}{ADS}]}.

\bibitem{K1976}
T.~W.~B. {Kibble}.
\newblock {Topology of cosmic domains and strings}.
\newblock {\em Journal of Physics A Mathematical General}, 9(8):1387--1398, Aug
  1976.
\newblock \href {http://dx.doi.org/10.1088/0305-4470/9/8/029} {\path{[DOI]}},
  {\small[\href{https://ui.adsabs.harvard.edu/abs/1976JPhA....9.1387K}{ADS}]}.

\bibitem{CK2010}
Edmund~J. {Copeland} and T.~W.~B. {Kibble}.
\newblock {Cosmic strings and superstrings}.
\newblock {\em Proceedings of the Royal Society of London Series A},
  466(2115):623--657, Mar 2010.
\newblock \href {http://arxiv.org/abs/0911.1345} {\path{arXiv:0911.1345}},
  \href {http://dx.doi.org/10.1098/rspa.2009.0591} {\path{[DOI]}},
  {\small[\href{https://ui.adsabs.harvard.edu/abs/2010RSPSA.466..623C}{ADS}]}.

\bibitem{BPRS2002}
F.~R. {Bouchet}, P.~{Peter}, A.~{Riazuelo}, and M.~{Sakellariadou}.
\newblock {Evidence against or for topological defects in the BOOMERanG data?}
\newblock {\em \prd}, 65(2):021301, Jan 2002.
\newblock \href {http://arxiv.org/abs/astro-ph/0005022}
  {\path{arXiv:astro-ph/0005022}}, \href
  {http://dx.doi.org/10.1103/PhysRevD.65.021301} {\path{[DOI]}},
  {\small[\href{https://ui.adsabs.harvard.edu/abs/2002PhRvD..65b1301B}{ADS}]}.

\bibitem{MP2014}
Adam {Moss} and Levon {Pogosian}.
\newblock {Did BICEP2 See Vector Modes? First B-Mode Constraints on Cosmic
  Defects}.
\newblock {\em \prl}, 112(17):171302, May 2014.
\newblock \href {http://arxiv.org/abs/1403.6105} {\path{arXiv:1403.6105}},
  \href {http://dx.doi.org/10.1103/PhysRevLett.112.171302} {\path{[DOI]}},
  {\small[\href{https://ui.adsabs.harvard.edu/abs/2014PhRvL.112q1302M}{ADS}]}.

\bibitem{W1985}
E.~{Witten}.
\newblock {Superconducting strings}.
\newblock {\em Nuclear Physics B}, 249:557--592, 1985.
\newblock \href {http://dx.doi.org/10.1016/0550-3213(85)90022-7}
  {\path{[DOI]}},
  {\small[\href{https://ui.adsabs.harvard.edu/abs/1985NuPhB.249..557W}{ADS}]}.

\bibitem{Slatyer:2009yq}
T.~R. {Slatyer}, N.~{Padmanabhan}, and D.~P. {Finkbeiner}.
\newblock {CMB constraints on WIMP annihilation: Energy absorption during the
  recombination epoch}.
\newblock {\em \prd}, 80(4):043526, August 2009.
\newblock \href {http://arxiv.org/abs/0906.1197} {\path{arXiv:0906.1197}},
  \href {http://dx.doi.org/10.1103/PhysRevD.80.043526} {\path{[DOI]}},
  {\small[\href{http://adsabs.harvard.edu/abs/2009PhRvD..80d3526S}{ADS}]}.

\bibitem{Kanzaki:2008qb}
T.~{Kanzaki} and M.~{Kawasaki}.
\newblock {Electron and photon energy deposition in the Universe}.
\newblock {\em \prd}, 78(10):103004, November 2008.
\newblock \href {http://arxiv.org/abs/0805.3969} {\path{arXiv:0805.3969}},
  \href {http://dx.doi.org/10.1103/PhysRevD.78.103004} {\path{[DOI]}},
  {\small[\href{http://adsabs.harvard.edu/abs/2008PhRvD..78j3004K}{ADS}]}.

\bibitem{Kanzaki:2009hf}
T.~{Kanzaki}, M.~{Kawasaki}, and K.~{Nakayama}.
\newblock {Effects of Dark Matter Annihilation on the Cosmic Microwave
  Background}.
\newblock {\em Progress of Theoretical Physics}, 123:853--865, May 2010.
\newblock \href {http://arxiv.org/abs/0907.3985} {\path{arXiv:0907.3985}},
  \href {http://dx.doi.org/10.1143/PTP.123.853} {\path{[DOI]}},
  {\small[\href{http://adsabs.harvard.edu/abs/2010PThPh.123..853K}{ADS}]}.

\bibitem{ZN1966}
Y.~B. {Zel'dovich} and I.~D. {Novikov}.
\newblock {The Hypothesis of Cores Retarded during Expansion and the Hot
  Cosmological Model}.
\newblock {\em \azh}, 43:758, 1966.
\newblock
  {\small[\href{https://ui.adsabs.harvard.edu/abs/1966AZh....43..758Z}{ADS}]}.

\bibitem{H1971}
S.~{Hawking}.
\newblock {Gravitationally collapsed objects of very low mass}.
\newblock {\em \mnras}, 152:75, 1971.
\newblock \href {http://dx.doi.org/10.1093/mnras/152.1.75} {\path{[DOI]}},
  {\small[\href{https://ui.adsabs.harvard.edu/abs/1971MNRAS.152...75H}{ADS}]}.

\bibitem{CH1974}
B.~J. {Carr} and S.~W. {Hawking}.
\newblock {Black holes in the early Universe}.
\newblock {\em \mnras}, 168:399--416, August 1974.
\newblock \href {http://dx.doi.org/10.1093/mnras/168.2.399} {\path{[DOI]}},
  {\small[\href{https://ui.adsabs.harvard.edu/abs/1974MNRAS.168..399C}{ADS}]}.

\bibitem{CKSY2010}
B.~J. {Carr}, Kazunori {Kohri}, Yuuiti {Sendouda}, and Jun'Ichi {Yokoyama}.
\newblock {New cosmological constraints on primordial black holes}.
\newblock {\em \prd}, 81:104019, May 2010.
\newblock \href {http://arxiv.org/abs/0912.5297} {\path{arXiv:0912.5297}},
  \href {http://dx.doi.org/10.1103/PhysRevD.81.104019} {\path{[DOI]}},
  {\small[\href{https://ui.adsabs.harvard.edu/\#abs/2010PhRvD..81j4019C}{ADS}]}.

\bibitem{P19761}
D.~N. {Page}.
\newblock {Particle emission rates from a black hole. II. Massless particles
  from a rotating hole}.
\newblock {\em \prd}, 14(12):3260--3273, Dec 1976.
\newblock \href {http://dx.doi.org/10.1103/PhysRevD.14.3260} {\path{[DOI]}},
  {\small[\href{https://ui.adsabs.harvard.edu/abs/1976PhRvD..14.3260P}{ADS}]}.

\bibitem{RW1957}
Tullio {Regge} and John~A. {Wheeler}.
\newblock {Stability of a Schwarzschild Singularity}.
\newblock {\em Physical Review}, 108(4):1063--1069, Nov 1957.
\newblock \href {http://dx.doi.org/10.1103/PhysRev.108.1063} {\path{[DOI]}},
  {\small[\href{https://ui.adsabs.harvard.edu/abs/1957PhRv..108.1063R}{ADS}]}.

\bibitem{MAC2016}
T.~N. {Ukwatta}, D.~R. {Stump}, J.~T. {Linnemann}, J.~H. {MacGibbon}, S.~S.
  {Marinelli}, T.~{Yapici}, and K.~{Tollefson}.
\newblock {Primordial Black Holes: Observational characteristics of the final
  evaporation}.
\newblock {\em Astroparticle Physics}, 80:90--114, Jul 2016.
\newblock \href {http://arxiv.org/abs/1510.04372} {\path{arXiv:1510.04372}},
  \href {http://dx.doi.org/10.1016/j.astropartphys.2016.03.007} {\path{[DOI]}},
  {\small[\href{https://ui.adsabs.harvard.edu/abs/2016APh....80...90U}{ADS}]}.

\bibitem{DM2010}
Marco {Cirelli}, Gennaro {Corcella}, Andi {Hektor}, Gert {H{\"u}tsi}, Mario
  {Kadastik}, Paolo {Panci}, Martti {Raidal}, Filippo {Sala}, and Alessandro
  {Strumia}.
\newblock {PPPC 4 DM ID: a poor particle physicist cookbook for dark matter
  indirect detection}.
\newblock {\em \jcap}, 2011(3):051, Mar 2011.
\newblock \href {http://arxiv.org/abs/1012.4515} {\path{arXiv:1012.4515}},
  \href {http://dx.doi.org/10.1088/1475-7516/2011/03/051} {\path{[DOI]}},
  {\small[\href{https://ui.adsabs.harvard.edu/abs/2011JCAP...03..051C}{ADS}]}.

\bibitem{PY2008}
Torbj{\"o}rn {Sj{\"o}strand}, Stephen {Mrenna}, and Peter {Skands}.
\newblock {A brief introduction to PYTHIA 8.1}.
\newblock {\em Computer Physics Communications}, 178(11):852--867, Jun 2008.
\newblock \href {http://arxiv.org/abs/0710.3820} {\path{arXiv:0710.3820}},
  \href {http://dx.doi.org/10.1016/j.cpc.2008.01.036} {\path{[DOI]}},
  {\small[\href{https://ui.adsabs.harvard.edu/abs/2008CoPhC.178..852S}{ADS}]}.

\bibitem{KY2000}
K.~{Kohri} and Jun'ichi {Yokoyama}.
\newblock {Primordial black holes and primordial nucleosynthesis: Effects of
  hadron injection from low mass holes}.
\newblock {\em \prd}, 61(2):023501, Jan 2000.
\newblock \href {http://arxiv.org/abs/astro-ph/9908160}
  {\path{arXiv:astro-ph/9908160}}, \href
  {http://dx.doi.org/10.1103/PhysRevD.61.023501} {\path{[DOI]}},
  {\small[\href{https://ui.adsabs.harvard.edu/abs/2000PhRvD..61b3501K}{ADS}]}.

\bibitem{GPR2017}
Alma~X. {Gonz{\'a}lez-Morales}, Stefano {Profumo}, and Javier
  {Reynoso-Cordova}.
\newblock {Prospects for indirect MeV dark matter detection with gamma rays in
  light of cosmic microwave background constraints}.
\newblock {\em \prd}, 96(6):063520, Sep 2017.
\newblock \href {http://arxiv.org/abs/1705.00777} {\path{arXiv:1705.00777}},
  \href {http://dx.doi.org/10.1103/PhysRevD.96.063520} {\path{[DOI]}},
  {\small[\href{https://ui.adsabs.harvard.edu/abs/2017PhRvD..96f3520G}{ADS}]}.

\bibitem{BGW2017}
Richard {Bartels}, Daniele {Gaggero}, and Christoph {Weniger}.
\newblock {Prospects for indirect dark matter searches with MeV photons}.
\newblock {\em \jcap}, 2017(5):001, May 2017.
\newblock \href {http://arxiv.org/abs/1703.02546} {\path{arXiv:1703.02546}},
  \href {http://dx.doi.org/10.1088/1475-7516/2017/05/001} {\path{[DOI]}},
  {\small[\href{https://ui.adsabs.harvard.edu/abs/2017JCAP...05..001B}{ADS}]}.

\bibitem{V1985}
A.~{Vilenkin}.
\newblock {Cosmic strings and domain walls.}
\newblock {\em \physrep}, 121:263--315, 1985.
\newblock \href {http://dx.doi.org/10.1016/0370-1573(85)90033-X}
  {\path{[DOI]}},
  {\small[\href{https://ui.adsabs.harvard.edu/abs/1985PhR...121..263V}{ADS}]}.

\bibitem{VS1994}
A.~{Vilenkin} and E.~P.~S. {Shellard}.
\newblock {\em {Cosmic strings and other topological defects}}.
\newblock Cambridge University Press, Cambridge, 1994.
\newblock
  {\small[\href{https://ui.adsabs.harvard.edu/abs/1994csot.book.....V}{ADS}]}.

\bibitem{OTW1986}
J.~P. {Ostriker}, C.~{Thompson}, and E.~{Witten}.
\newblock {Cosmological effects of superconducting strings}.
\newblock {\em Physics Letters B}, 180:231--239, November 1986.
\newblock \href {http://dx.doi.org/10.1016/0370-2693(86)90301-1}
  {\path{[DOI]}},
  {\small[\href{https://ui.adsabs.harvard.edu/abs/1986PhLB..180..231O}{ADS}]}.

\bibitem{VV1987}
A.~{Vilenkin} and T.~{Vachaspati}.
\newblock {Electromagnetic radiation from superconducting cosmic strings}.
\newblock {\em Physical Review Letters}, 58:1041--1044, March 1987.
\newblock \href {http://dx.doi.org/10.1103/PhysRevLett.58.1041} {\path{[DOI]}},
  {\small[\href{https://ui.adsabs.harvard.edu/abs/1987PhRvL..58.1041V}{ADS}]}.

\bibitem{CSV2012}
Yi-Fu Cai, Eray Sabancilar, and Tanmay Vachaspati.
\newblock {Radio bursts from superconducting strings}.
\newblock {\em Phys. Rev.}, D85:023530, 2012.
\newblock \href {http://arxiv.org/abs/1110.1631} {\path{arXiv:1110.1631}},
  \href {http://dx.doi.org/10.1103/PhysRevD.85.023530} {\path{[DOI]}}.

\bibitem{CSSDV2012}
Yi-Fu {Cai}, Eray {Sabancilar}, Dani{\`e}le~A. {Steer}, and Tanmay
  {Vachaspati}.
\newblock {Radio broadcasts from superconducting strings}.
\newblock {\em \prd}, 86(4):043521, Aug 2012.
\newblock \href {http://arxiv.org/abs/1205.3170} {\path{arXiv:1205.3170}},
  \href {http://dx.doi.org/10.1103/PhysRevD.86.043521} {\path{[DOI]}},
  {\small[\href{https://ui.adsabs.harvard.edu/abs/2012PhRvD..86d3521C}{ADS}]}.

\bibitem{VV1985}
T.~{Vachaspati} and A.~{Vilenkin}.
\newblock {Gravitational radiation from cosmic strings}.
\newblock {\em \prd}, 31:3052--3058, June 1985.
\newblock \href {http://dx.doi.org/10.1103/PhysRevD.31.3052} {\path{[DOI]}},
  {\small[\href{https://ui.adsabs.harvard.edu/abs/1985PhRvD..31.3052V}{ADS}]}.

\bibitem{MN2013}
Koichi {Miyamoto} and Kazunori {Nakayama}.
\newblock {Cosmological and astrophysical constraints on superconducting cosmic
  strings}.
\newblock {\em \jcap}, 2013(7):012, Jul 2013.
\newblock \href {http://arxiv.org/abs/1212.6687} {\path{arXiv:1212.6687}},
  \href {http://dx.doi.org/10.1088/1475-7516/2013/07/012} {\path{[DOI]}},
  {\small[\href{https://ui.adsabs.harvard.edu/abs/2013JCAP...07..012M}{ADS}]}.

\bibitem{KT1982}
T.~W.~B. {Kibble} and Neil {Turok}.
\newblock {Self-intersection of cosmic strings}.
\newblock {\em Physics Letters B}, 116(2-3):141--143, Oct 1982.
\newblock \href {http://dx.doi.org/10.1016/0370-2693(82)90993-5}
  {\path{[DOI]}},
  {\small[\href{https://ui.adsabs.harvard.edu/abs/1982PhLB..116..141K}{ADS}]}.

\bibitem{C2015}
Jens {Chluba}.
\newblock {Green's function of the cosmological thermalization problem - II.
  Effect of photon injection and constraints}.
\newblock {\em \mnras}, 454(4):4182--4196, Dec 2015.
\newblock \href {http://arxiv.org/abs/1506.06582} {\path{arXiv:1506.06582}},
  \href {http://dx.doi.org/10.1093/mnras/stv2243} {\path{[DOI]}},
  {\small[\href{https://ui.adsabs.harvard.edu/abs/2015MNRAS.454.4182C}{ADS}]}.

\end{thebibliography}
\end{document}